\documentclass[12pt]{article}
\usepackage{amsmath,amsfonts,amsbsy}
\usepackage[dvips]{graphicx,epsfig}
\usepackage[vcentermath,enableskew]{youngtab}

\def\hybrid{\topmargin -20pt    \oddsidemargin 0pt
        \headheight 0pt \headsep 0pt
        \textwidth 6.25in       
        \textheight 9.5in       
        \marginparwidth .875in
        \parskip 5pt plus 1pt   \jot = 1.5ex}

\hybrid

\renewcommand{\theequation}{\thesection.\arabic{equation}} \csname
@addtoreset\endcsname{equation}{section}

\newcommand{\hoch}[1]{$^{#1}$}
\newcommand{\bein}[3]{{#1}_{#2}^{\hspace{0.5em}{#3}}}

\def\moth{\mathsurround=0pt}
\newdimen\zo \zo=0pt

\def\tick{\leaders\hrule height 0.5ex depth 0pt \hskip 0.5pt}
\def\upboxfill{$\moth \setbox\zo\hbox{\tick}%
  \hskip 3pt\hbox to 0pt{$\tick$\hss}\hrulefill \hbox to 7.5pt{$\tick$\hss}$}
\def\underbox#1{\offinterlineskip{\mathord{\mathop{\vtop{\moth\ialign{##\crcr
      $\hfil\displaystyle{#1}\hfil$\crcr\noalign{}
      {\upboxfill}\crcr\noalign{}}}}\limits}}}
\def\dtick{\leaders\hrule height .34pt depth 0.5ex \hskip 0.5pt}
\def\downboxfill{$\moth \setbox\zo\hbox{\dtick}%
  \hskip 2pt\hbox to 0pt{$\dtick$\hss}\hrulefill \hbox to 2pt{$\dtick$\hss}$}

\def\undersym#1{\underbox{{}#1}}

\def\bec{\begin{center}}
\def\ec{\end{center}}

\def\c{\gamma} 

\def\d{\delta} 

\def\e{\epsilon}

\def\l{\lambda}
\def\L{\Lambda}
\def\m{\mu}
\def\n{\nu}
\def\r{\rho}
\def\s{\sigma}

\def\t{\tau}

\def\O{\Omega}
\def\om{\omega}

\def\cF{{\cal F}}

\def\cA{{\cal A}}

\def\cW{{\cal W}}

\def\cA{{\cal A}}

\def\del{\partial}
\def\na{\nabla}
\let\la=\label

\def\nn{\nonumber}
\newcommand{\eq}[1]{(\ref{#1})}

\def\be{\begin{equation}}
\def\ee{\end{equation}}
\def\bea{\begin{eqnarray}}
\def\eea{\end{eqnarray}}
\def\ba{\begin{array}}
\def\ea{\end{array}}

\def\ft#1#2{{\textstyle{{\scriptstyle #1}
\over {\scriptstyle #2}}}}

\def\scs#1{\section{\sc #1}}
\def\scss#1{\subsection{\sc #1}}

\begin{document}

\hfill{\texttt{ITP-UU-07/40}}

\vspace{-5pt}

\hfill{\texttt{SPIN-07/28}}

\vspace{-5pt}


\hfill{August 2007}

\vspace{20pt}

\begin{center}


{\Large\sc Geometry and Dynamics\\[0.2cm] of Higher-Spin Frame Fields}


\vspace{30pt}
{\sc Johan Engquist\hoch1 and  Olaf Hohm\hoch1}\\[15pt]

\hoch{1}{\it\small Institute for Theoretical Physics and Spinoza
Institute, Utrecht University \\3508 TD Utrecht, The Netherlands}


\vspace{35pt}  \end{center}

\begin{center} {\bf ABSTRACT}\\[3ex]

\begin{minipage}{13cm}
\small We give a systematic account of unconstrained free bosonic
higher-spin fields on $D$-dimensional Minkowski and (Anti-)de
Sitter spaces in the frame formalism. The generalized spin
connections are determined by solving a chain of torsion-like
constraints. Via a generalization of the vielbein postulate these
allow to determine higher-spin Christoffel symbols, whose relation
to the de\hspace{0.1cm}Wit--Freedman connections is discussed. We
prove that the generalized Einstein equations, despite being of
higher-derivative order, give rise to the AdS Fr\o nsdal equations
in the compensator formulation. To this end we derive Damour-Deser
identities for arbitrary spin on AdS. Finally we discuss the
possibility of a geometrical and local action principle, which is
manifestly invariant under unconstrained higher-spin symmetries.

\end{minipage}
\end{center}

 {\vfill\leftline{}\vfill \vskip  10pt \footnoterule
{\footnotesize E-mails: \texttt{J.Engquist@phys.uu.nl,
O.Hohm@phys.uu.nl} \vskip -12pt}}

\setcounter{page}{1}

\pagebreak

\tableofcontents

\scs{Introduction} In the ongoing search for the underlying
principles and symmetries of string/M-theory, the analysis of
higher-spin (HS) theories has played a distinguished role
\cite{Gross:1988ue,Isberg:1993av,Sundborg:2000wp,Sezgin:2002rt,
Sagnotti:2003qa,Engquist:2005yt,Bonelli:2003kh,Fotopoulos:2007nm}.
Since the original work of Fr\o nsdal
\cite{Fronsdal:1978rb,Fronsdal:1978vb} in the 1970's there has
been substantial progress in the following two directions. On the
one hand, due to the work of Vasiliev, the problem of constructing
consistently interacting HS theories coupled to gravity has been
efficiently attacked
\cite{Vasiliev:1990en,Vasiliev:1990vu,Vasiliev:2003ev} (see also
\cite{Bekaert:2005vh} and references therein). On the other hand,
the free higher-spin theories on Minkowski and AdS have been
recently reformulated in a much more geometrical fashion, using
`unconstrained fields' (a term to be defined below), leading to
manifestly HS invariant Einstein-like equations
\cite{Francia:2002aa,Francia:2002pt,Bekaert:2003az,Bekaert:2006ix}.

However, there is a certain tension between these two directions in
the way they treat the HS fields. The approach of Vasiliev to
construct interacting HS theories is based on the gauging of certain
HS algebras in a similar fashion to gravity and supergravity
theories. Like in supergravity this requires a frame-like
formulation of HS fields, which generalizes the vielbein formalism
of general relativity. Such a formalism has been developed by
Vasiliev in \cite{Vasiliev:1980as}. (For a related approach see
\cite{Aragone:1981yn,Aragone:1988yx}.) It describes a spin-$s$ field
by a 1-form $\bein{e}{\mu}{a_1\cdots a_{s-1}}$ carrying $s-1$
totally symmetric frame indices, together with a generalized spin
connection $\om_\m{}^{a_1\cdots a_{s-1},b}$. In contrast, the recent
progress in the geometrical formulation of free HS theories is based
on a `metric-like' formulation, in which a spin-$s$ field is
described by totally symmetric covariant tensor
$h_{\mu_1\cdots\mu_s}$ of rank $s$, generalizing the metric tensor
of Einstein gravity. However, in the latter approach it has been
clear since the early work of de\hspace{0.1cm}Wit and Freedman
\cite{deWit:1979pe} that the natural object, replacing the
Christoffel symbol of Riemannian geometry, is not a single
connection but instead a hierarchy of $s-1$ connections, each being
expressed as derivatives of the previous one. The top connection is
identified with the HS Riemann tensor, which in turn is an
$s$-derivative object of the physical spin-$s$ field.

Until recently it has, accordingly, been unclear, how to take
advantage of this appealing geometrical structure in that there
seems to be no natural way to get standard 2$^{\rm nd}$ order
field equations. In order to recover the 2$^{\rm nd}$ order
spin-$s$ equations formulated by Fr\o nsdal one has to work with
`constrained' objects, in which the transformation parameter of
the HS gauge symmetry is restricted to be traceless. This has the
consequence that the trace of the 2$^{\rm nd}$
de\hspace{0.1cm}Wit--Freedman connection is invariant under the
constrained HS symmetry. Consequently, it gives rise to the Fr\o
nsdal equations without employing the higher connections. Though
this invariance under constrained HS transformations is sufficient
to decouple the unphysical longitudinal degrees of freedom of a
massless spin-$s$ state, the constrained formulation seems to be
unnatural once the coupling to gravity is considered since then,
ultimately, the metric and thus the operation of `taking the
trace' becomes dynamical. Therefore an unconstrained formulation
is clearly desirable.

Progress into this direction has been achieved due to the work of
Francia and Sagnotti \cite{Francia:2002aa,Francia:2002pt}. (Apart
from that, unconstrained HS fields appeared in
\cite{Collins:1987bd,Bandos:2005mb,Buchbinder:2007ak,Buchbinder:2006eq,Fotopoulos:2007yq,
de Medeiros:2003dc}, but also in string field theory, see
\cite{Bengtsson:1986ys,Sagnotti:2003qa,Bekaert:2003uc} and
references therein.) Their formulation is entirely geometrical in
that it is given in terms of the trace-full, i.e.~unconstrained, HS
Riemann tensor. It requires, however, \textit{non-local} field
equations, in which inverse powers of the Laplacian appear in order
to get rid of the higher derivative order. On the other hand, there
exists an unconstrained local formulation, which handles the
non-invariance of the Fr\o nsdal equations under trace-full HS
transformations via introducing so-called compensator fields
\cite{Francia:2005bu,Francia:2007qt}. Unfortunately, this
formulation is not geometrical anymore in the sense that the
compensators are unrelated to the connections or the curvature
tensor. However, it has been shown by Bekaert and Boulanger in
\cite{Bekaert:2003az,Bekaert:2006ix}, based on \cite{Damour:1987vm},
how to obtain in flat space a theory which is, first, purely
geometrical and, second, local but still equivalent to the 2$^{\rm
nd}$ order Fr\o nsdal equations. For this they proved that the
generalized Einstein equations can be locally integrated via the
Poincar\'e lemma, giving rise to the 2$^{\rm nd}$ order compensator
formulation of \cite{Francia:2002aa,Francia:2002pt}, with the
compensator fields appearing as `integration constants'.

So on the one hand, the most geometrical formulation for
metric-like fields seems to require the full chain of
de\hspace{0.1em}Wit--Freedman connections and higher-derivative
field equations. On the other hand, the frame-like formulation for
free HS fields possesses only a single spin-connection-like
object. However, in HS gauge theories (based on the associated HS
algebras) higher connections do appear. These are the so-called
`extra fields'. But usually they are treated on a different
footing than the lowest connection, since they can be decoupled at
the free field level \cite{Fradkin:1986qy,Vasiliev:2001wa}. This
is usually done precisely to guarantee 2$^{\rm nd}$ order free
field equations. But this is only possible if one works with
trace-less tensors (i.e.~with $SO(D)$ instead of $GL(D)$ Young
tableaux), in accordance with the fact that the higher
de\hspace{0.1em}Wit--Freedman connections are only required in the
trace-full, i.e.~unconstrained case.

An extension of the Vasiliev formulation to unconstrained objects
has been initiated in \cite{Sagnotti:2005ns}. It has been shown
for a spin-3 field on Minkowski space that also in this formalism
the Fr\o nsdal equations can be obtained from geometrical
higher-derivative equations. More recently, this spin-3 analysis
has been extended to AdS in \cite{Engquist:2007kz}. However, in
spite of these partial results (see also \cite{Bekaert:2006ix}), a
complete account of an unconstrained frame formalism for arbitrary
spin on AdS backgrounds, generalizing that of
\cite{Vasiliev:1980as}, is so far lacking. Instead the more
general, but also more abstract technique of the so-called
$\sigma^{-}$-- cohomology has been developed, which allows to
analyze the dynamical content of HS theories to a large extent
without requiring explicit formulas of the type we are going to
consider in this paper (for an introduction in the unconstrained
case see \cite{Bekaert:2005vh}). We think, however, that a more
explicit treatment will be crucial for concrete applications. This
leads us to reexamine the geometry of HS frame fields in a
systematic and self-contained fashion.

More specifically, the paper is organized as follows. In sec.~2 we
review linearized gravity and the metric-like formulation of free HS
fields in flat space together with the de\hspace{0.1em}Wit--Freedman
connections. The frame-like formalism of HS geometry will be
discussed in sec.~3. We derive the torsion constraints, solve them
explicitly and discuss the resulting Bianchi identities. Finally, we
discuss the relation to the de\hspace{0.1em}Wit--Freedman
connections. In sec.~4 we analyze the dynamics of HS fields, and we
show that the Einstein equations give rise to the compensator form
of the Fr\o nsdal equations. For that purpose we derive a
generalization of the Damour-Deser identity to AdS. We close with a
brief discussion concerning the possibility of an action principle.
Our conventions, the technical details related to the solutions of
the torsion constraints and the proof of the Damour-Deser identity
are contained in the appendices A--C.

\scs{Linearized gravity and higher-spin fields} \scss{Christoffel
and spin connection for spin-2}\label{spin2rev} One way to think
about HS theories is as a generalization of Einstein gravity, in
which the spin-2 metric tensor is replaced by a higher rank
tensor. So, in order to set the stage for our later examinations,
let us first recall the spin-2 case.

The Riemannian geometry underlying Einstein's theory can be
formulated either in terms of the metric $g_{\mu\nu}$ or a frame
field (`vielbein') $e_\mu{}^a$. Since we will later on focus on HS
fields that generalize the frame formulation of gravity, we start
by reviewing the vielbein formalism. At the purely kinematical
level this formalism allows a gauge theoretical interpretation in
the sense that the vielbein and the spin connection are viewed as
components of a gauge connection, which is associated to the
(Anti-)de Sitter or Poincar\'e algebra.\footnote{However, as far
as the dynamics is concerned, this interpretation breaks down,
except in special cases like three-dimensional gravity
\cite{Witten:1988hc} or specific forms of Lovelock gravities in
odd dimensions \cite{Chamseddine:1989nu,Chamseddine:1990gk}, which
can be viewed as Yang-Mills gauge theories based on Chern-Simons
forms. In connection to HS and Kaluza-Klein theories see
\cite{Engquist:2007kz} and \cite{Hohm:2005sc}, respectively.} We
will focus on the AdS case, but keep the cosmological constant
explicit such that the In\"on\"u-Wigner contraction
$\Lambda\rightarrow 0$ can always be performed. The Lie algebra of
the required AdS group $SO(D-1,2)$ in $D$ dimensions is spanned by
generators $M_{AB}=-M_{BA}$ and reads
 \begin{eqnarray} \label{so42alg}
 \begin{split}
  [M_{AB},M_{CD}]&\ =\
  \eta_{BC}M_{AD}-\eta_{AC}M_{BD}-\eta_{BD}M_{AC}+\eta_{AD}M_{BC}
   \;\\ &\ \equiv \ f_{AB,CD}{}^{EF}M_{EF}\ ,
 \end{split}
\end{eqnarray}
where
 \bea \la{structco}
  \eta_{AB}&=&{\rm diag}(-1,1,1,1,1,-1 )\ , \qquad
  f_{AB,CD}{}^{EF}\;=\;4\d_{[A}^{[E} \eta_{B][C}^{} \d^{F]}_{D]} \ .
 \eea
(For our conventions see appendix A.) Next we need to split this
basis into a Lorentz covariant form, i.e.
 \bea
  \begin{split}
   [M_{ab},M_{cd}]&\ =\
   \eta_{bc}M_{ad}-\eta_{ac}M_{bd}-\eta_{bd}M_{ac}+\eta_{ad}M_{bc}\ ,
   \cr [M_{ab},P_c]&\ =\ -2\eta_{c[a}P_{b]}\ , \qquad
   [P_a,P_b]\;=\;\L M_{ab}\ ,
  \end{split}
 \eea
where we have defined $P_a=\sqrt \L M_{a0'}=M_{a0'}/L$, with the
AdS length $L$. Then writing the associated gauge field as
$A_{\mu}=\bein{e}{\mu}{a}P_a+\ft12\bein{\omega}{\mu}{ab}M_{ab}$,
we obtain the required identification of vielbein and spin
connection. Moreover, the torsion and curvature tensors naturally
appear as the non-abelian field strengths of (\ref{so42alg}).
Explicitly, one has
 \bea
  F\ =\ dA+A\wedge A\  =\ T^aP_a+\ft12{\cal R}^{ab}M_{ab}\;,
 \eea
which indeed reads
 \begin{eqnarray}\label{torcur}
  T^a&=& D^{L}e^a \ \equiv\ de^a+\omega^a{}_b\wedge e^b\;, \\
  {\cal R}^{ab}&=&R^{ab}+\Lambda e^a\wedge e^b\ \equiv \
  d\omega^{ab}+\omega^{ac}\wedge \omega_c{}^b+\Lambda e^a\wedge e^b  \;.
 \end{eqnarray}
Imposing the condition $T^a=0$ allows to solve for the spin
connection $\bein{\omega}{\mu}{ab}$ in terms of derivatives of the
vielbein. Inserting this into $R_{\mu\nu}{}^{ab}$ gives rise to
the standard Riemann tensor as a function of second derivatives of
the metric.

Let us now turn to the metric formulation, in which the
Levi-Civita connection is encoded in the Christoffel symbols. To
see where these enter, we first note that the Lorentz covariant
derivative $D_{\mu}^{L}\bein{e}{\nu}{a}$ of the vielbein is not
covariant under diffeomorphisms, in contrast to the antisymmetric
part $D_{[\mu}^{L}\bein{e}{\nu]}{a}$, which is the torsion 2-form
defined in (\ref{torcur}). So in order to define a derivative
which is covariant with respect to local Lorentz transformations
and diffeomorphisms, we have to introduce a connection which is
symmetric in $\mu,\nu$. These are the Christoffel symbols, related
to the spin connection via the metricity condition (`vielbein
postulate')
 \bea\label{nonlinmet}
  D_{\mu}e_\nu{}^a\ =\ \partial_{\mu}\bein{e}{\nu}{a}+\bein{\omega}{\mu}{ab}e_{\nu b}-
  \Gamma_{\mu\nu}^{\rho}\bein{e}{\rho}{a} \ = \ 0\;.
 \eea

In order to compare with the free HS dynamics to be discussed
below, we have to consider linearized gravity. By linearizing
around Minkowski space,
$\bein{e}{\mu}{a}=\bein{\delta}{\mu}{a}+\kappa\bein{h}{\mu}{a}$,
the metricity condition (\ref{nonlinmet}) reads at the first
non-trivial order
 \bea\label{metricity0}
  \partial_{\mu}h_{\nu \rho}+\omega_{\mu|\rho,\nu}
  -\Gamma_{\rho,\mu\nu}\ =\ 0\;,
 \eea
in which the distinction between flat and curved indices is now
redundant and will therefore henceforth be ignored.\footnote{Here,
we use the notation $\omega_{\mu|\rho,\nu}$, in which a comma
separates indices in different rows of a Young-tableau, in order
to comply with the notation used for HS fields later on.} At this
stage, the frame formulation has the basic consequence that
$h_{\mu \nu}$ is not symmetric, but also has an antisymmetric
part. Accompanied with this is an additional gauge symmetry,
namely the local Lorentz symmetry. In the linearization the latter
reads together with the diffeomorphisms
 \bea\label{linlor}
  \delta_{\xi}h_{\mu \nu} \ =\  \partial_{\mu}\xi_{\nu}-\Lambda_{\nu,\mu}\;, \qquad
 \eea
i.e.~the Lorentz transformations act as St\"uckelberg symmetries
with anti-symmetric shift parameter $\Lambda_{\mu,\nu}$, while the
connections transform as
 \bea
  \begin{split}
   \delta_{\Lambda}\omega_{\mu|\nu,\rho} &=
   \partial_{\mu}\Lambda_{\nu,\rho}\;, \qquad
   \delta_{\xi}\omega_{\mu|\nu,\rho}\ =\ 0\;, \\
   \delta_{\xi}\Gamma_{\rho,\mu\nu} &=
   \partial_{\mu}\partial_{\nu}\xi_{\rho}\;, \qquad
   \delta_{\Lambda}\Gamma_{\rho,\mu\nu} \ =\ 0\;.
  \end{split}
 \eea
Using the explicit expression for the spin connection obtained
from (\ref{torcur}),
 \bea
  \omega_{\mu|\nu,\rho}\ =\ \partial_{\mu}h_{[\nu\rho]}
  +\partial_{\rho}h_{(\mu\nu)}-\partial_{\nu}h_{(\mu\rho)}\;,
 \eea
which depends also on the antisymmetric part, one derives from
(\ref{metricity0}) for the Christoffel symbols\footnote{The reader
might miss a factor of $\ft12$, which is due to the chosen
normalization in the expansion of $e_\mu{}^a$ around flat space.}
 \bea\label{christ}
  \Gamma_{\rho,\mu\nu} \ =\
  \partial_{\mu}h_{(\rho\nu)}+\partial_{\nu}h_{(\rho\mu)}-\partial_\rho
  h_{(\mu\nu)}\;.
 \eea
Note that here the antisymmetric part drops out, in agreement with
the fact that $\Gamma_{\rho,\mu\nu}$ is Lorentz gauge-invariant.
These shift-symmetries can in turn be used to fix a gauge, in which
the anti-symmetric part of $h_{\mu\nu}$ is gauged away and only the
symmetric part survives. As this gauge-fixing would be violated by
generic diffeomorphisms $\xi^{\rho}$, this requires a compensating
Lorentz transformation with parameter
$\Lambda_{\nu,\mu}=\partial_{[\mu}\xi_{\nu]}$, giving rise to the
standard diffeomorphism symmetry on $h_{\mu\nu}$,
 \bea
  \delta_{\xi}h_{\mu\nu}\ =\ \partial_{(\mu}\xi_{\nu)}\;, \qquad
  \delta_{\xi}\omega_{\mu|\nu,\rho}\ =\ \del_{\mu}\del_{[\rho}\xi_{\nu]}\;.
 \eea
Here, after gauge-fixing, also the spin connection transforms
non-trivially under diffeomorphisms.

\scss{Higher-spin fields and de\hspace{0.1em}Wit--Freedman
connections}\label{dewit} After reviewing the spin-2 case, we will
now discuss HS fields. These are given by totally symmetric tensor
fields $h_{\mu_1\cdots\mu_s}$ of rank $s$, which generalize the
metric tensor of Einstein gravity. In the massless case their
dynamics has to permit a gauge symmetry, which eliminates the
unphysical longitudinal degrees of freedom and generalizes the
diffeomorphism symmetry of general relativity. It is parameterized
by a symmetric rank $s-1$ tensor $\epsilon_{\mu_1\cdots\mu_{s-1}}$
and reads
 \bea\label{HSsym}
  \delta_{\epsilon}h_{\mu_1\cdots\mu_s} \ =\
  \partial_{(\mu_1}\epsilon_{\mu_2\cdots\mu_s)}\;.
 \eea
An action which generalizes the linearized Einstein-Hilbert term
to higher spin and stays invariant under (\ref{HSsym}) has been
given by Fr\o nsdal \cite{Fronsdal:1978rb}. It can be written as
 \bea
   \nn S_{\rm F}[h] &=& \frac{1}{2}\int d^D x
   \big(\partial_{\mu}h_{\nu_1\cdots\nu_s}\partial^{\mu}h^{\nu_1\cdots\nu_s}
   -\ft12s(s-1)\partial_{\mu}h^{\prime}_{\nu_3\cdots\nu_s}\partial^{\mu}
   h^{\prime\nu_3\cdots\nu_s}
 \eea
 \vspace{-1.1cm}
 \bea\label{Fronsdal}
   &&\qquad \qquad \qquad\qquad \quad+s(s-1)\partial_{\mu}h^{\prime}_{\nu_3\cdots\nu_s}
   \partial\cdot h^{\mu\nu_3\cdots\nu_s}-s\partial\cdot
   h_{\nu_2\cdots\nu_s}\partial\cdot h^{\nu_2\cdots\nu_s} \\
   \nn&&\qquad \qquad \qquad\qquad \quad-\ft14 s(s-1)(s-2)\partial\cdot
   h^{\prime}_{\nu_1\cdots\nu_{s-3}}\partial\cdot h^{\prime
   \nu_1\cdots\nu_{s-3}}\big)\;.
 \eea
Here $h^{\prime}$ denotes the trace in the Minkowski metric and
$\partial\cdot
h_{\nu_2\cdots\nu_s}=\partial^{\rho}h_{\rho\nu_2\cdots\nu_2}$. The
action (\ref{Fronsdal}) is invariant under (\ref{HSsym}), provided
the transformation parameter is traceless. Moreover, the double
trace of $h$ has to be set to zero in order for the field to
describe the correct number of degrees of freedom
\cite{deWit:1979pe}. This is the so-called constrained
formulation, for which in total
 \bea\label{constraint0}
  \epsilon^{\prime}_{\hspace{0.1cm}\nu_3\cdots\nu_{s-1}}\ =\ \epsilon^\rho{}_{\rho\nu_3\cdots\nu_{s-1}}\ = \ 0\;, \qquad
  h^{\prime\prime}_{\hspace{0.1cm}\rho_5\cdots\rho_s}\ =\ h^{\mu\nu}{}_{\mu\nu\rho_5\cdots\rho_s}\ =\ 0\;.
 \eea
Note that the double-tracelessness constraint stays invariant
under the constrained HS symmetries. The spin-$s$ field equations
derived from \eq{Fronsdal} read
 \bea
  {\cal F}_{\mu_1\cdots \mu_s}\ = \ \square h_{\mu_1\cdots \mu_s}
  -s\partial_{(\mu_1}\partial\cdot
  h_{\mu_2\cdots\mu_s)}+\frac{s(s-1)}{2}\partial_{(\mu_1}^{}
  \partial_{\mu_2}^{}h_{\mu_3\cdots\mu_s)}^{\prime}\ =\ 0\;,
 \eea
which defines the Fr\o nsdal operator ${\cal F}$.

This formulation of HS dynamics can be extended to AdS
backgrounds. First, the minimal substitution
$\partial_{\mu}\rightarrow \nabla_{\mu}$ in the action
(\ref{Fronsdal}) and the symmetry variations (\ref{HSsym}), with
$\nabla_{\mu}$ denoting the AdS covariant derivative, violates the
HS invariance. For maximally symmetric backgrounds this can be
compensated by adding a mass-like term proportional to the
cosmological constant \cite{Fronsdal:1978vb}. Then the equations
of motion are
 \bea
  {\cal F}^{\rm AdS}_{\mu_1\cdots \mu_s}& = & \square h_{\mu_1\cdots \mu_s}
  -s\nabla_{(\mu_1}\nabla\cdot
  h_{\mu_2\cdots\mu_s)}+\frac{s(s-1)}{2}\nabla_{(\mu_1}^{}
  \nabla_{\mu_2}^{}h_{\mu_3\cdots\mu_s)}^{\prime}\\ \nn &&-\Lambda\Big(\big((D-3+s)(s-2)-s\big)h_{\mu_1\cdots \mu_s}+
  s(s-1)g_{(\m_{1}\m_{2}}^{}h'_{\mu_3\cdots \mu_s)}\Big)\ = \ 0\;.
 \eea
Its variation under unconstrained HS transformations reads
 \bea
  \la{gvoF} \delta_{\epsilon}{\cal F}_{\mu_1\cdots \mu_s}^{\rm AdS}\
  =\  \frac{(s-1)(s-2)}{2}
  \left(\nabla_{(\mu_1}\nabla_{\mu_2}\nabla_{\mu_3}
  \epsilon^{\prime}_{\mu_4\cdots\mu_s)}-4\Lambda g_{(\mu_1\mu_2}\nabla_{\mu_3}
  \epsilon^{\prime}_{\mu_4\cdots\mu_s)}\right)\;,
 \eea
and so, indeed, it is only invariant if $\epsilon^{\prime}=0$.

The Fr\o nsdal action (\ref{Fronsdal}) together with the
constraints (\ref{constraint0}) consistently describes the free
propagation of a massless spin-$s$ field. However, as it stands,
the formulation is not very geometrical, since (\ref{Fronsdal})
has been determined by hand, and there is no obvious way to
rewrite it in terms of HS curvatures. This is in contrast to the
spin-2 case, for which (\ref{Fronsdal}) coincides with the
linearized Einstein-Hilbert action, and can therefore be written
in a manifestly spin-2, that is, diffeomorphism invariant form. A
first step towards a more geometrical formulation of HS fields
would accordingly require a generalization of the Christoffel
symbols of Riemannian geometry in an unconstrained way. An
appealing formalism has been presented already by
de\hspace{0.1cm}Wit and Freedman in \cite{deWit:1979pe} for the
flat case, and will be reviewed in the following. (It has only
been possible recently to tackle the AdS analogue, at least
perturbatively in the inverse AdS radius \cite{Manvelyan:2007hv}.)

To start with, we have to find Christoffel symbols, which are
first order in derivatives of the HS fields and transform as
connections. The definition\footnote{Here we have chosen a
different overall sign than in \cite{deWit:1979pe}.}
 \bea\label{christ1}
  \Gamma^{(1)}_{\rho,\mu_1\cdots\mu_s}\ =\ -\partial_{\rho}h_{\mu_1\cdots\mu_s}
  +s\partial_{(\mu_1}h_{\mu_2\cdots\mu_s)\rho}\;,
 \eea
gives rise to the transformation behaviour
 \bea\label{christ1trans}
  \delta_{\epsilon}\Gamma^{\tiny (1)}_{\rho,\mu_1\cdots\mu_s}
  \ =\ (s-1)
  \partial_{(\mu_1}\partial_{\mu_2}\epsilon_{\mu_3\cdots\mu_s)\rho}\;.
 \eea
This transformation is as simple as possible, since the relative
coefficients in (\ref{christ1}) have been chosen such that the index
$\rho$ appears in (\ref{christ1trans}) only on the transformation
parameter. Moreover, for spin-2 it reduces to the standard
Christoffel symbol (\ref{christ}). However, the transformation
(\ref{christ1trans}) is not truly connection-like, since it does not
allow for the definition of an invariant curvature tensor in terms
of derivatives of $\Gamma^{(1)}_{\rho,\mu_1\cdots\mu_s}$. Instead,
it allows to define a hierarchy of $s$ connection-like objects, each
of which being defined as derivatives of the previous one via the
recursion relation ($1\leq m\leq s$)
 \bea\label{deWconn}
  \Gamma^{(m)}_{\rho_1\cdots\rho_m,\mu_1\cdots\mu_s}\ =\
  \partial_{\rho_1}^{}
  \Gamma^{(m-1)}_{\rho_2\cdots\rho_m,\mu_1\cdots\mu_s}-\frac{s}{m}
  \partial_{(\mu_1}^{}\Gamma^{(m-1)}_{|\rho_2\cdots\rho_m,\rho_1|
  \mu_2\cdots\mu_s)}\;.
 \eea
These are totally symmetric in the two sets of indices $\mu_i$ and
$\rho_i$, although for the latter it is not manifest. They
transform under the HS symmetry as
 \bea\label{convar}
  \delta_{\epsilon}\Gamma_{\rho_1\cdots\rho_m,\mu_1\cdots\mu_s}^{(m)}\
  = \ (-1)^{m+1} {s-1 \choose m}
  \del_{(\mu_1}\cdots\del_{\mu_{m+1}}\epsilon_{\mu_{m+2}\cdots\mu_s)\rho_1\cdots\rho_m}\;.
 \eea
Therefore only the $(s-1)$-connection transforms as a proper
connection in the sense that
 \bea
  \delta_{\epsilon}\Gamma^{(s-1)}_{\rho_1\cdots\rho_{s-1},\mu_1\cdots\mu_s}
  \ =\ (-1)^{s} \partial_{\mu_1}\cdots\del_{\mu_s}\epsilon_{\rho_1\cdots\rho_{s-1}}\;.
 \eea
Consequently, the top `connection' in (\ref{deWconn}) is actually
an invariant curvature tensor defined in terms of derivatives of
$\Gamma^{(s-1)}$,
 \bea\label{riemann}
  R_{\rho_1\cdots\rho_s,\mu_1\cdots\mu_s}\ =\
  \Gamma^{(s)}_{\rho_1\cdots\rho_s,\mu_1\cdots\mu_s}\;.
 \eea

So there exists a geometrical structure for unconstrained HS
fields, which naturally extends the known spin-1 and spin-2 cases.
In particular, the gauge-invariant curvatures are $s$-derivative
objects, generalizing the 1$^{\rm st}$ order field strength of
electrodynamics and the 2$^{\rm nd}$ order Riemann tensor of
general relativity. However, due to this higher-derivative nature
it is not clear how to obtain sensible 2$^{\rm nd}$ order field
equations. This is where the constrained formulation enters. In
fact, once we constrain as in (\ref{constraint0}), the trace of
the second de\hspace{0.1em}Wit--Freedman connection $\Gamma^{(2)}$
turns out to be invariant, since it transforms only into the trace
part of $\epsilon$, as can be seen from (\ref{convar}). Invariant
2$^{\rm nd}$ order field equations can then be written as
 \bea
  \Gamma^{(2)\rho}{}_{\rho,\mu_1\cdots\mu_s}\ =\ 0\;,
 \eea
which coincide with the Fr\o nsdal equations -- as it should be by
gauge invariance. More recently it has been shown that the
generalized Einstein equation, stating that the trace of the HS
Riemann tensor (\ref{riemann}) vanishes, effectively also gives
rise to 2$^{\rm nd}$ order field equations through local
integrations \cite{Bekaert:2003az,Bekaert:2006ix}. This analysis
will be extended to the frame formalism and AdS later on.

\scs{Higher-spin geometry}\label{HSalg} In this section we start
the analysis of HS fields in the frame formalism. We introduce the
corresponding Lie algebra, which determines the HS gauge
symmetries, and whose field strengths will be interpreted as HS
torsion and curvature tensors, respectively. After imposing
torsion constraints we give their explicit solutions and discuss
the resulting Bianchi identities together with the relation to the
de\hspace{0.1em}Wit--Freedman connections.

 \scss{Higher-spin gauge algebra and
connections} We are going to consider the free dynamics of HS
fields on AdS. Therefore, we have to fix the spin-2 vielbein to
the background value $\bar{e}_\mu{}^a$ of the AdS geometry, being
covariantly characterized by $\bar{{\cal R}}^{ab}=0$. In order to
introduce the HS frame fields, we extend the Lie algebra
$\frak{so}(D-1,2)$ in (\ref{so42alg}) for each spin-$s$ field by a
generator $Q_{A_1\cdots A_{s-1},B_1\cdots B_{s-1}}$, which lives
in the two-row Young tableaux
 \bea
  Q_{A(s-1),B(s-1)}: \qquad \underbrace{\yng(5,5)\cdots\yng(5,5)}_{s-1}\
  ~~  .
 \eea
Here and in the following we use the convenient short-hand
notation $A(s)=A_1\cdots A_s$ for totally symmetric indices and
similarly for Lorentz indices. The commutation relations with the
spin-2 or $\frak{so}(D-1,2)$ generators are entirely fixed by
representation theory and read
 \bea \la{spintwos}
  \nn [M_{AB},Q_{C(s-1),D(s-1)}]&=&-4(s-1)\eta\undersym{{}_{A\langle C_{s-1}}Q_{|B|}}\hspace{0cm}{}_{C(s-2),D(s-1)\rangle}
  \\ & = &
  \nonumber
  -2\Big(\eta\undersym{{}_{AC_1}Q_{\phantom{|}B\phantom{|}}}\hspace{-.1cm}{}_{C_2\cdots
  C_{s-1},D(s-1)}
  +\eta\undersym{{}_{A\phantom{\langle}\hspace{-.1cm}C_2}Q_{C_1B\hspace{.1cm}}}{}_{\hspace{-.1cm}C_3\cdots C_{s-1},D(s-1)}
  +\cdots \\&&
  \qquad+\eta\undersym{{}_{AD_{s-1}}Q_{C(s-1),D_1\cdots
  D_{s-2}}{}_B\hspace{.1cm}}\Big)\;.
 \eea
Here we introduced projectors according to the Young symmetries
imposed by the left-hand side. (For our conventions see appendix
A.) As in the spin-2 case above we are next splitting into a
Lorentz covariant basis, for which the generators
$Q_{A(s-1),B(s-1)}$ decompose into $Q_{a(s-1),b(t)}$ for $0\le
t\le s-1$, being in the Young tableau

\vspace{-3.0cm}

  \setlength{\unitlength}{0.8cm}
  \begin{picture}(10,5)
  \put(2.45,-0.13){$Q_{a(s-1),b(t)}$:}\put(13.0,-0.13){.}
  \put(7.55,-0.13){$\overbrace{\yng(8,5)}^{s-1}$}
    \put(7.55,-0.45){$\underbrace{\phantom{\yng(5)}}_{t}$}
  \end{picture} \vspace{1cm} \

\vspace{.0cm}

\noindent The commutation relations \eq{spintwos} then read
 \bea \la{lorentz}
    \begin{split} [M_{ab},Q_{c(s-1),d(s-1)}]&\ =\
        -4(s-1)\eta\undersym{{}_{a\langle
        c_{s-1}}Q_{|b|\hspace{.07cm}}}\hspace{-0.07cm}{}_{c(s-2),d(s-1)\rangle} \ , \\
        [M_{ab},Q_{c(s-1),d(t)}]&\ =\ -2\big((s-1)\eta\undersym{{}_{a\langle
        c_{s-1}}Q_{|b|\hspace{.07cm}}}\hspace{-0.07cm}{}_{c(s-2),d(t)\rangle}+t\eta\undersym{{}_{a\langle
        d_t}Q_{c(s-1),|b|\hspace{.07cm}}}\hspace{-0.07cm}{}_{d(t-1)\rangle}\big)\ , \\
        [P_a,Q_{c(s-1),d(t)}]&\ =\ t(s-t+1)\frac{s-1}{s-t}\eta_{a\langle c_{s-1}}Q_{c(s-2)d_t,d(t-1)\rangle}\\
        &\qquad -(s-t-1) \L Q_{c(s-1),d(t)a} \ ,
     \end{split}
 \eea
where the brackets $\langle\hspace{0.1em},\rangle$ impose the the
corresponding Young projections. To keep track of the powers of
$\L$, dimensional analysis is useful. By defining the (mass)
dimensions $[\L]=[L^{-2}]=2$ and $[M_{AB}]=0$ it follows that
$[P_a]=1$ and $[Q_{a(s-1),b(t)}]=s-1-t$. The Lie brackets $[Q,Q]$
vanish, since these would correspond to self-interactions of the
HS fields, which do not enter the free dynamics. In fact, the
algebra \eq{spintwos} spanned by $M_{AB}$ and the $Q$ can be
viewed as a truncation of the infinite-dimensional HS algebras
(sometimes denoted by $\frak{ho}(D-1,2)$) of Vasiliev
\cite{Fradkin:1986ka,Bekaert:2005vh}, which appear upon
linearization \cite{Sagnotti:2005ns,Engquist:2007kz}.

Next we are introducing a HS gauge field $\cA=\cA_\m dx^\m$, which
reads in components
 \bea\label{gaugeexp}
  {\cal A}_{\mu}\ = \ \bein{\bar{e}}{\mu}{a}P_a +
  \frac{1}{2}\bein{\bar{\omega}}{\mu}{ab}M_{ab} + \sum_{s=3}^\infty\cW_\m^{(s)}\;,
 \eea
where the spin-$s$ contribution is given by
  \bea
   \cW_\m^{(s)}& =&
   \frac{1}{(s-1)!}e_\mu{}^{a(s-1)}Q_{a(s-1)}+\sum_{t=1}^{s-1}\frac{s-t}{s!t!}\om_\mu{}^{a(s-1),b(t)}Q_{a(s-1),b(t)}\ .
  \eea
The coefficients $(s-t)/(s!t!)$ impose unit-strength
normalizations and follow from the Hook length formula (see,
e.g.,~\cite{Fuchs:1997jv}) for an $(s-1,t)$ Young tableau. The
fields $\om_\mu{}^{a(s-1),b(t)}$ will be referred to in the
following as HS connections, while the fields $e_\m{}^{a(s-1)}$
will later on be identified with the physical spin-$s$ fields, or
the generalized vielbeins. We define the canonical dimension
$[\cA]=1$, from which it follows that $[e_\m{}^{a(s-1)}]=2-s$ and
$[\om_\mu{}^{a(s-1),b(t)}]=2-s+t$. In order to derive HS torsions
and curvatures we compute as above the non-abelian HS field
strength, which is given by
\begin{eqnarray}
  {\cal F}_{\mu\nu} \ =\  \partial_{\mu}{\cal W}_{\nu} -\partial_{\nu}{\cal W}_{\mu}
  +[{\cal W}_{\mu},{\cal W}_{\nu}] \ = \ \bar{T}_{\mu\nu}{}^{a}P_a + \frac{1}{2}\bar{{\cal
  R}}_{\mu\nu}{}^{ab}M_{ab}+ \sum_{s=3}^\infty\cF_{\m\n}^{(s)}\ ,
\end{eqnarray}
where the spin-$s$ curvatures read
\begin{eqnarray}
   \la{rt99}
   \cF_{\m\n}^{(s)}\ = \ \sum_{t=0}^{s-2}\frac{s-t}{s!t!}T_{\m\n}{}^{a(s-1),b(t)}Q_{a(s-1),b(t)}
   +\frac{1}{s!(s-1)!}R_{\mu\nu}{}^{a(s-1),b(s-1)}Q_{a(s-1),b(s-1)}\   .
\end{eqnarray}
By using \eq{lorentz} we find
\begin{eqnarray} \label{strength}
    \begin{split}
        R_{\m\n}{}^{a(s-1),b(s-1)}& \ = \ \bar
        D_{\m}\om_\n{}^{a(s-1),b(s-1)}-2(s-1) \L\om_\n{}^{\langle
        a(s-1),b(s-2)} \bar e_\m{}^{b_{s-1}\rangle}- (\mu \leftrightarrow \n)\ , \\
        T_{\m\n}{}^{a(s-1),b(t)}& \ = \ \bar
        D_{\m}\om_\n{}^{a(s-1),b(t)}-\om_\n{}^{
        a(s-1),b(t)c}\bar e_{\m c}\\ &\qquad -t(s-t+1)\L\om_\n{}^{\langle
        a(s-1),b(t-1)}\bar e_\m{}^{b_t\rangle}- (\mu \leftrightarrow \n)\ ,
    \end{split}
\end{eqnarray}
where it is understood that $\om_\m{}^{a(s-1)}\equiv
e_\m{}^{a(s-1)}$. Here we defined Lorentz covariant derivatives,
which are given by
 \bea
    \nn \bar D_{\m}\om_\n{}^{a(s-1),b(t)}& = &
    \del_{\m}\om_\n{}^{a(s-1),b(t)}+(s-1)\bar
    \om_\m^{\langle a_{s-1}|a|}\om_{\n a}{}^{a(s-2),b(t)\rangle}\\
    &&+t\hspace{.07cm}\bar\om_\m{}^{\langle
    b_t|a|}\om_{\n}{}^{a(s-1),}{}_a{}^{b(t-1)\rangle}\;.
 \eea

Finally, we give the HS gauge transformations, under which the
field strengths above stay invariant. Defining the transformation
parameter to be
 \bea
  \epsilon \ = \ \frac{1}{(s-1)!}\epsilon^{a(s-1)}Q_{a(s-1)}
  +\sum_{t=1}^{s-t}\frac{s-1}{s!t!}\epsilon^{a(s-1),b(t)}Q_{a(s-1),b(t)}\;,
 \eea
the gauge variation $\delta {\cal W}_{\mu} = D_{\mu}\epsilon =
\partial_{\mu}\epsilon+[{\cal W}_{\mu},\epsilon]$ of the
various components reads
  \bea \la{gaugetr}
    \begin{split}
        \d_\e e_\m{}^{a(s-1)} & \ =\  \bar D_\m \e^{a(s-1)} -\e^{
        a(s-1),c}\bar e_{\m c}\ , \\
        \d_\e \om_\m{}^{a(s-1),b(t)} &\ = \ \bar D_{\m}\e^{a(s-1),b(t)}-\e^{
        a(s-1),b(t)c}\bar e_{\m c}-t(s-t+1)\L\e^{\langle
        a(s-1),b(t-1)}\bar e_\m{}^{b_t\rangle}\ .
    \end{split}
 \eea
The parameter $\epsilon^{a(s-1)}$ corresponding to the lowest HS
generator will later give rise to the `physical' HS symmetry. In
contrast, the higher symmetries given by $\epsilon^{a(s-1),b(t)}$
for $t\geq 1$ act as St\"uckelberg shift symmetries, that correspond
to the linearized Lorentz transformations in (\ref{linlor}) and
which subsequently will be gauge-fixed.

\scss{Torsion constraints and their solutions}\label{torsection}
We now turn to the torsion constraints for the HS connections. As
in the spin-2 case reviewed in sec.~\ref{spin2rev}, we identify
the field strengths associated to the lowest gauge fields
(containing the physical HS field) with the torsion tensors, while
the top component of the field strength will be identified with
the HS generalization of the Riemannian curvature tensor, cf.~the
expansion in \eq{rt99}.

Specifically, we impose the constraints
 \bea\label{constraint}
  T_{\m\n}{}^{a(s-1),b(t)} \ = \ 0 \ , \qquad 0\le t\le s-2\ \;.
 \eea
These allow to solve for the connection
$\bein{\omega}{\mu}{a(s-1),b(t)}$ at level $t$ in terms of the
first derivative of the `previous' connection
$\bein{\omega}{\mu}{a(s-1),b(t-1)}$. (This parallels the
elimination of the so-called `extra fields' in terms of the
physical field carried out in the constrained formalism in
\cite{Vasiliev:1986td,Lopatin:1987hz}.) To this end we introduce
HS coefficients of anholonomity, which read
 \bea \la{omcdab1} \begin{split} \O^{bc|a(s-1)}& =  \bar e^{\m
  b}\bar e^{\n c}\big(\bar
  D_\m e_\n{}^{a(s-1)}  -\bar D_\n e_\m{}^{a(s-1)}\big)\ , \\ \la{omcdab2}
  \O^{cd|a(s-1),b(t)}& =  \bar e^{\m c}\bar e^{\n d}\big(\bar D_\m
  \om_\n{}^{a(s-1),b(t)}+\Lambda t(s-t+1)\omega_\mu{}^{\langle
  a(s-1),b(t-1)}\bar{e}_\nu{}^{b_t\rangle} -(\mu\leftrightarrow\nu
  )\big)\  . \end{split}
 \eea
The torsion constraints \eq{constraint} can then be written as
 \bea \O^{cd|a(s-1),b(t)}\ = \
  \om^{d|a(s-1),b(t)c}-\om^{c|a(s-1),b(t)d}\ ,
 \eea
where we converted the 1-form index on $\om$ into a flat one. We
start with the first constraint in (\ref{constraint}), i.e.~$t=0$,
in order to determine the HS connection at $t=1$. One finds the
general solution
 \bea\label{firstsol}
  \omega^{a|b(s-1),c}\ =\ \frac{1}{2}\left(\Omega^{ca|b(s-1)}+\Omega^{c(b_1|b_2\cdots
  b_{s-1})a}+\Omega^{a(b_1|b_2\cdots b_{s-1})c}\right)+\xi^{b(s-1),ac}
  \;.
 \eea
Here $\xi^{b(s-1),ac}$ denotes a gauge degree of freedom. To be
more precise, we note that for (\ref{firstsol}) to transform as a
HS connection according to (\ref{gaugetr}), we have to assign a
non-trivial transformation behaviour to $\xi$:
 \bea\label{compsym}
  \delta_{\epsilon}\xi_{b(s-1),ac}\ =\ \hspace{0.1em}\bar{D}_{\langle
  a}\epsilon_{b(s-1),c\rangle}-s\hspace{0.2em}\epsilon_{\langle
  b(s-1)}\eta_{ac\rangle}-\epsilon_{b(s-1),ac}\;.
 \eea
In particular, we see that it is subject to the St\"uckelberg
shift symmetry parametrized by $\epsilon_{b(s-1),ac}$. Therefore,
by gauge-fixing this symmetry, $\xi$ can be set to zero. In terms
of the HS connection $\omega$ this can be interpreted as follows.
A priori, $\omega$ takes values in the Young tableaux
 \bea\label{omegadeco}
  \yng(1)\otimes \overbrace{\yng(5,1)}^{s-1}\ = \ \overbrace{\yng(5,1,1)}^{s-1}\ \oplus\
  \overbrace{\yng(6,1)}^s \  \oplus \ \overbrace{\yng(5,2)}^{s-1}\;.
 \eea
The part in the $(s-1,2)$ tableau is precisely given by
$\xi^{b(s-1),ac}$ and is not determined by the torsion constraint.
As such it can be gauge-fixed to zero. Note that by \eq{compsym}
this requires a compensating gauge transformation with
$\epsilon_{b(s-1),ac}=\bar{D}_{\langle
  a}\epsilon_{b(s-1),c\rangle}-s\hspace{0.2em}\epsilon_{\langle
  b(s-1)}\eta_{ac\rangle}$.

Before we turn to the higher connections we are going to perform
another gauge-fixing. In fact, so far the HS field
$\bein{e}{\mu}{a(s-1)}$ carries irreducible parts according to the
decomposition
 \bea
    \yng(1)\ \otimes \ \underbrace{\yng(2)\cdots \yng(2)}_{s-1}\ = \ \underbrace{\yng(2)\cdots\yng(3)}_s\ \oplus\
    \overbrace{\yng(5,1)}^{s-1}\ \;,
 \eea
i.e.~it contains the completely symmetric physical part
 \bea
  h_{\mu_1\cdots\mu_s}:=\bein{\bar{e}}{(\mu_1}{a_1}\cdots\bein{\bar{e}}{\mu_{s-1}}{a_{s-1}}
  e_{\mu_s)a_1\cdots a_{s-1}}^{}\;,
 \eea
but also the hook-like part. The latter can in turn be gauged away
by fixing the symmetry spanned by $\epsilon_{a(s-1),b}$,
cf.~\eq{gaugetr}. The residual gauge symmetry on $h$ is then given
by
 \bea
  \delta_{\epsilon}h_{\mu_1\cdots\mu_s} =
  \nabla_{(\mu_1}\epsilon_{\mu_2\cdots\mu_s)}\;,
 \eea
in agreement with \eq{HSsym}, but for AdS
backgrounds.\footnote{Here and in the following we denote the
AdS-background covariant derivative acting only on curved indices
by $\nabla_{\mu}$.} After gauge-fixing, the first HS connection
can be rewritten as
 \bea
  \omega^{a|b(s-1),c}\ = \ \frac{s}{s-1} \mathbb{P}_{(s-1,1)} \Omega^{ca|b(s-1)}\;.
 \eea
The resulting connection then takes values only in the $(s,1)$
tableaux. For instance, in case of a spin-3 field on flat space
one can show

\vspace{-2.7cm}

  \setlength{\unitlength}{0.8cm}
  \begin{picture}(6,5)
  \put(1,0.5){\qquad\quad $\om_{\m|\n(2),\r}\ = \
  \Bigg(\frac12\Big({\mathbb P}_{\footnotesize{\yng(3,1)}}+{\mathbb
  P}_{\footnotesize{\yng(3,1)}}\Big)-{\mathbb
  P}_{\footnotesize{\yng(3,1)}}\Bigg)\del_\m h_{\r\n(2)}$}
  \put(6.7,0.34){$\m$} \put(7.17,.33){$\n_1$} \put(7.67,.34){$\r$}
  \put(6.64,-.17){$\n_2$} \put(9.18,0.34){$\m$}
  \put(9.61,.33){$\n_2$} \put(10.18,.34){$\r$}
  \put(9.14,-.17){$\n_1$} \put(11.96,0.34){$\m$}
  \put(12.41,.33){$\n_1$} \put(12.91,.34){$\n_2$}
  \put(11.96,-.17){$\r$} \put(15.91,.35){.} \end{picture} \vspace{1cm} \

\vspace{-.6cm}


It turns out that, in order to solve the torsion constraints for
$t>1$, first the lower constraints have to be solved. 
This requirement comes about as follows: Provided the background
geometry is AdS, the torsion constraint at level $t$ implies a
Bianchi identity, which in turn gives a condition on the HS
coefficients of anholonomity. In the next section we will prove that
this reads
 \bea
  \Omega^{[ab|c(s-1),d(t-1)|e]} = 0 \;.
 \eea
Moreover, after gauge-fixing this relation extends further in that
antisymmetrization of $a,b$ with any of the remaining indices
gives zero. This allows to prove that the following expressions
solve the torsion constraints:
 \bea\label{Omgsol}
  \om_{\m|\n(s-1),\r(t)}\ = \ \frac{s-t+1}{s-t}\mathbb
  P_{(s-1,t)}\O_{\r_t\m|\n(s-1),\r(t-1)}+\xi_{\nu(s-1),\rho(t)\mu}\;.
 \eea
Here $\xi$ denotes again a gauge-degree of freedom. To be more
precise, in analogy to (\ref{omegadeco}), the irreducible parts of
$\omega^{(s-1,t)}$ read
 \bea\label{decorep}
  (1) \ \otimes \ (s-1,t) \ = \ (s,t) \ \oplus \ (s-1,t,1) \
  \oplus \
  (s-1,t+1)\;,
 \eea
and $\xi$ is the part in the $(s-1,t+1)$ Young tableau. It
corresponds to the shift symmetry in this tableau, which is given by
$\delta_{\epsilon}\xi_{a(s-1),b(t+1)}=-\epsilon_{a(s-1),b(t+1)}$. In
order to solve for the HS connections at level $t+1$ in terms of the
coefficients of anholonomity at level $t$, also this symmetry has to
be gauge-fixed. This gives rise to the compensating transformation
 \bea
  \epsilon_{\nu(s-1),\rho(t+1)}\ =\
  \nabla_{\langle\rho_1}\cdots\nabla_{\rho_{t+1}}
  \epsilon_{\nu(s-1)\rangle}^{} +{\cal O}(\L)\;,
 \eea
which in turn expresses all transformation parameter in terms of
the physical HS symmetry. After this gauge-fixing, only the
$(s,t)$ contribution in the decomposition (\ref{decorep}) is
non-zero. This follows from the fact that the antisymmetrization
over three indices in $(s-1,t,1)$ vanishes identically due to the
total symmetry of $h$ and of the partial derivatives (see
eq.~\eq{flatconn} below).

To better understand (\ref{Omgsol}) we first examine it on flat
space, where it can be rewritten in a closed form in terms of
derivatives of the physical HS fields:
 \bea\label{flatconn}
  \om_{\m|\n(s-1),\r(t)}\ = \ \frac{s}{s-t}\mathbb
  P_{(s-1,t)}\del_{\r(t)}h_{\n(s-1)\m}\ = \
  \frac{s}{s-t}\del_{\langle\r(t)}h_{\n(s-1)\rangle\m}\;.
 \eea
Writing out the projector explicitly one finds for $t=1$
 \bea
  \om_{\m|\n(s-1),\r}& = & \del_{\r} h_{\n(s-1)\m}-
  \del_{(\n_1} h_{\n_2\cdots\nu_{s-1})\r\m}\;,
 \eea
while the higher ones are given by
\bea
\om_{\m|\n(s-1),\r(t)}& = & \mathbb P_{\n(s-1)}{\mathbb
P}_{\r(t)}\sum_{m=0}^t{t \choose m}(-1)^m\del_{\r(t-m)\n(m)}
h_{\n_{m+1}\cdots\n_{s-1}\r_{t-m+1}\cdots\r_{t}\m}\ , \eea where
$\mathbb{P}_{\nu(s-1)}$, etc.~imposes unit-strength
symmetrization.

Let us now inspect the AdS case more explicitly. Even though we have
a closed expression in terms of the $\Omega$ in (\ref{Omgsol}), it
turns out to be much more tedious to give these in terms of
(AdS-covariant) derivatives of the physical HS fields. While the
first connection ($t=1$) coincides with the flat space expression
upon a minimal substitution,
 \bea \la{eqxx}
  \omega_{\mu|\nu(s-1),\rho}\ =\ \nabla_{\rho}h_{\nu(s-1)\mu}
  -\nabla_{(\nu_1}h_{\nu_2\cdots\nu_{s-1})\mu}\;,
 \eea
this is not so for $t\geq 2$. Indeed, there are correction terms
proportional to $\Lambda$, which read for $t=2$
 \bea \la{t2solu} \om_{\m|\n(s-1),\r(2)}& = &
  \frac{s}{s-2}\nabla_{\langle\r(2)}h_{\n(s-1)\rangle\m}-\L\frac{s(s-1)}{s-2}g_{\langle\r(2)}h_{\n(s-1)\rangle\m}\;.
 \eea
In order to determine analogous expressions for all higher
connections, one can prove a recursion relation by use of
\eq{Omgsol}, whose partial solutions we give in appendix
\ref{appendixRec}. In total we have
\bea
     \nn\om_{\m|\n(s-1),\r(t)}& = &
     \frac{s}{s-t}\nabla_{\langle\r_1}\cdots\nabla_{\r_t}h_{\n(s-1)\rangle\m}\\ \la{omeads9} &&
     +\sum_{k=1}^{[t/2]}\L^k\c_{k,t}g_{\langle\r_1\r_2}\cdots
     g_{\r_{2k-1}\r_{2k}}\nabla_{\r_{2k+1}}\cdots\nabla_{\r_t}h_{\n(s-1)\rangle\m}\ . \eea
Here the brackets $[\hspace{0.25em}]$ denote the largest integer
smaller or equal to the argument. The coefficients $\gamma_{k,t}$
are determined by eq.~(\ref{xfr}) in appendix \ref{appendixRec}.

\scss{Bianchi identities}\la{sectionbianchi} After having solved
the torsion constraints, we are now going to discuss some of its
consequences, specifically the generalized Bianchi identities. In
form language (\ref{strength}) reads
 \begin{eqnarray}\nonumber
  T^{a(s-1),b(t)} &=&
  \bar{D}\omega^{a(s-1),b(t)}+t(s-t+1)\Lambda\omega^{\langle
  a(s-1),b(t-1)}\wedge\bar{e}^{b_t\rangle}\\
  && +\omega^{a(s-1),b(t)c}\wedge\bar{e}_c \ =\ 0 \;,
  \\ \nonumber
  R^{a(s-1),b(s-1)} &=&
  \bar{D}\omega^{a(s-1),b(s-1)}+2(s-1)\Lambda\omega^{\langle
  a(s-1),b(s-2)}\wedge\bar{e}^{b_{s-1}\rangle}\;.
 \end{eqnarray}
Let us assume that we have solved the first $t-1$ torsion
constraints. By acting with $\bar{D}$ on $T^{a(s-1),b(t-1)}$ we
can conclude
 \bea
  0 \ =\  \bar{D}T^{a(s-1),b(t-1)} \ =\
  T^{a(s-1),b(t-1)c}\wedge\bar{e}_c\;,
 \eea
where we also used $T^{a(s-1),b(t-2)}=0$. This implies that prior
to solving the torsion constraint at level $t$, the Bianchi
identities for the previous ones yield already a nontrivial
relation. In components it reads
 \bea
  T_{[\mu\nu |\rho(s-1),\sigma(t-1)|\lambda]}\ = \ 0\;,
 \eea
where all indices are curved. Rewriting $T$ in terms of $\Omega$
and using the symmetries of the HS connections in the frame
indices, one concludes
 \bea \la{omegid}
  \Omega^{[ab|c(s-1),d(t-1)|e]} \ =\  0 \;,
 \eea
which was required in order to solve the torsion constraint at
level $t$. Finally, applying $\bar{D}$ to the last torsion
constraint $t=s-2$, we derive similarly an algebraic Bianchi
identity for the Riemann tensor,
 \bea\label{algbi}
  R_{[\mu\nu|\rho]\sigma(s-2),\lambda(s-1)}\ = \ 0\;.
 \eea

Let us now turn to the differential Bianchi identities of the
Riemann tensor. Acting again with $\bar{D}$ and using the AdS
relation $\bar{R}^{ab}=-\Lambda\bar{e}^a\wedge\bar{e}^b$ we conclude
 \bea
  \bar{D}R^{a(s-1),b(s-1)}\ =\ 0\;,
 \eea
which reads in components with curved incdices
 \bea\label{diffbianchi}
  \nabla_{\mu}R_{\nu\rho|\sigma(s-1),\lambda(s-1)}
  +\nabla_{\nu}R_{\rho\mu|\sigma(s-1),\lambda(s-1)}
  +\nabla_{\rho}R_{\mu\nu|\sigma(s-1),\lambda(s-1)}\ =\ 0\;.
 \eea
Finally this gives rise to a contracted Bianchi identity:
Introducing a Ricci tensor
 \bea
  ({\rm Ric})_{\mu\nu|\rho(s-1),\sigma(s-3)}\ =\
  R_{\mu\nu|\rho(s-1),\sigma(s-3)\hspace{0.5em}\lambda}^{\hspace{5.7em}\lambda}\;,
 \eea
and an analogue of the curvature `scalar'
 \bea
  S_{\mu|\nu(s-2),\rho(s-3)}\ = \ g^{\sigma\lambda}({\rm
  Ric})_{\sigma\mu|\lambda\nu(s-2),\rho(s-3)}\;,
 \eea
one can define the `Einstein tensor'
 \bea
  G_{\nu\rho|\mu\sigma(s-2),\lambda(s-3)}\ = \ ({\rm
  Ric})_{\nu\rho|\mu\sigma(s-2),\lambda(s-3)}-
  2g_{\mu[\nu}S_{\rho]|\sigma(s-2),\lambda(s-3)}\;.
 \eea
This has been defined such that it satisfied the conservation law
 \bea
  \nabla^{\mu}G_{\nu\rho|\mu\sigma(s-2),\lambda(s-3)} \ =\  0\;,
 \eea
which follows by contracting (\ref{diffbianchi}) with
$g^{\mu\sigma_1}$ and $g^{\lambda_{s-2}\lambda_{s-1}}$.

 \scss{Generalized metricity condition}\label{metricity}
Let us now relate the HS connections determined above to the
de\hspace{0.1cm}Wit--Freedman connections reviewed in
sec.~\ref{dewit}. For this we first introduce generalized
Christoffel symbols in complete analogy to the spin-2 case (see
sec.~\ref{spin2rev}). More specifically, we want to add connection
terms to the partial derivative
$\partial_{\mu}\bein{e}{\nu}{a_1\cdots a_{s-1}}$ such that the
resulting expression is HS invariant. We first note that for the
antisymmetric part in $\mu,\nu$ we know already the answer, namely
the required connections are precisely the HS connections
$\bein{\omega}{\mu}{a_1\cdots a_{s-1},b}$ above, as these combine
to the HS invariant torsion tensors (\ref{strength}). To define
the symmetric part as well, we simply introduce as in the spin-2
case a symmetric Christoffel symbol which does the job. Explicitly
we define $\widehat{\Gamma}_{\rho(s-1),\mu\nu}$ through the
generalized metricity condition
 \bea\label{metricitycond}
  {\cal
  D}_{\mu}h_{\nu\rho(s-1)}\ =\ \partial_{\mu}h_{\nu\rho(s-1)}+\omega_{\mu|\rho(s-1),\nu}
  -\widehat{\Gamma}^{(1)}_{\rho(s-1)|\mu\nu}\ =\ 0\;,
 \eea
which defines a HS covariant derivative ${\cal D}_{\mu}$. These
Christoffel symbols have a different index structure than the
de\hspace{0.1em}Wit--Freedman connections, and accordingly they
have a different transformation behaviour. The latter is
determined by (\ref{metricitycond}) as follows
 \bea
  \delta_{\epsilon}\widehat{\Gamma}^{(1)}_{\rho(s-1)|\mu\nu} = \frac{2}{s}
  \partial_{\mu}\partial_{\nu}\epsilon_{\rho(s-1)}+\frac{2(s-2)}{s}\partial_{(\mu}
  \partial_{(\rho_1}\epsilon_{\rho_2\cdots\rho_{s-1})\nu)}
  -\frac{s-2}{s}\partial_{(\rho_1}\partial_{\rho_2}\epsilon_{\rho_3\cdots\rho_{s-1})\mu\nu}\;.
 \eea
This allows to express these connections in terms of the
de\hspace{0.1em}Wit--Freedman connections as
 \bea
  \widehat{\Gamma}^{(1)}_{\rho(s-1)|\mu\nu}\ =\ \Gamma^{(1)}_{(\rho_1,\rho_2\cdots\rho_{s-1})\mu\nu}
  -\frac{s-2}{2(s-1)}\left(\Gamma^{(1)}_{\mu,\rho(s-1)\nu}+\Gamma^{(1)}_{\nu,\rho(s-1)\mu}\right)\;,
 \eea
which can be verified by explicit evaluation. We see that both
types of generalized Christoffel connections coincide for spin
$s=2$. $\widehat{\Gamma}$ can be written in a closed form as
 \bea
  \widehat{\Gamma}_{\rho(s-1)|\mu\nu}\ =\ \partial_{\mu}h_{\nu\rho(s-1)}
  +\partial_{\nu}h_{\rho(s-1)\mu}-\partial_{(\rho_1}h_{\rho_2\cdots\rho_{s-1})\mu\nu}\;,
 \eea
which is the obvious generalization of the spin-2 expression
\eq{christ}.

The metricity relation can be directly extended to all higher
connections and for AdS, by imposing
 \bea
  {\cal D}_{\mu}\omega_{\nu|\rho(s-1),\sigma(t)} \ = \
  \nabla_{\mu}\omega_{\nu|\rho(s-1),\sigma(t)}+\omega_{\mu|\rho(s-1),\sigma(t)\nu}
  -\widehat{\Gamma}_{\rho(s-1),\sigma(t)|\mu\nu}^{(t)} \ = \
  0\;.
 \eea
After gauge-fixing the physical HS field $h_{\mu_1\cdots\mu_s}$ to
the completely symmetric part, these equations define a hierarchy
of generalized Christoffel symbols, which are metric-like
$t$-derivative objects in $h$. Nevertheless they are different
from the de\hspace{0.1em}Wit--Freedman connections, though they
can be expressed in terms of them. However, since these relations
are not relevant for our present analysis, we refrain from
computing them here explicitly for general $t$.

\scs{Higher-spin dynamics} In this section we will analyze the free
HS dynamics in the geometrical frame formalism developed so far.
Specifically, we will show that the Einstein-like equations, stating
the vanishing of the HS Ricci tensor, are equivalent to the Fr\o
nsdal equations in the unconstrained compensator formulation for
Minkowski space as well as for AdS. Finally, we discuss the
possibility of a geometrical action principle.

\scss{Generalized Einstein equations on Minkowski} We are going to
prove that the HS Einstein equation
 \bea\label{HSEinstein}
  ({\rm Ric})_{\mu\nu|\rho(s-1),\sigma(s-3)}\ =\
  R_{\mu\nu|\rho(s-1),\sigma(s-3)\hspace{0.5em}\lambda}^{\hspace{5.7em}\lambda}=0\;,
 \eea
is equivalent to the Fr\o nsdal formulation. The equation
(\ref{HSEinstein}) appears in the unconstrained Vasiliev equations
\cite{Sagnotti:2005ns}, but more recently also through a
Chern-Simons action principle \cite{Engquist:2007kz}. The former
equations are first order differential equations in the so-called
unfolded formulation (see \cite{Vasiliev:2005zu,Vasiliev:1992gr}
and \cite{Barnich:2004cr}) and read in the linearization
\cite{Sagnotti:2005ns,Bekaert:2005vh}
 \bea\label{Vasiliev}
  {\cal F}^{a(s-1),b(t)}\ = \ \delta_{t,s-1}\hspace{0.2em}\bar{e}_c
  \wedge\bar{e}_d\hspace{0.2em} C^{a(s-1)c,b(s-1)d}\;.
 \eea
Here ${\cal F}$ denotes the (trace-full) HS field strength defined
in sec.~\ref{HSalg} and $C^{a(s),b(s)}$ is the so-called Weyl
zero-form. The latter is traceless and so (\ref{Vasiliev}) implies
the dynamical equation (\ref{HSEinstein}) together with the
torsion constraints discussed in sec.~\ref{torsection}. The same
dynamical equation has been derived from the Chern-Simons action
principle of \cite{Engquist:2007kz}, upon imposing the torsion
constraints. (A more detailed discussion about the possibility of
an action principle giving rise to (\ref{HSEinstein}) will be
presented below.)

To explain the approach, we start with the low spin cases $s=3$
and $s=4$, improving the discussion of
\cite{Sagnotti:2005ns,Engquist:2007kz}. The explicit form of the
spin-3 connections can be read off from (\ref{flatconn}).
Inserting into the HS Riemann tensor implies
 \bea\label{vanishcurl}
  ({\rm Ric})_{\mu\nu|\rho\sigma}\ = \ 2\partial_{[\mu}{\cal F}_{\nu]\rho\sigma} \
  =\
  0\;,
 \eea
where the first equation follows by explicit evaluation. This is
the so-called Damour-Deser identity, which has first been derived
for spin-3 geometry in metric-like formulation in
\cite{Damour:1987vm}.\footnote{For a flat-space generalization to
fields in mixed Young tableaux see \cite{Bekaert:2006ix}.} This
relation in turn allows to locally integrate by virtue of the
Poincar\'e lemma, resulting in ${\cal F}_{\mu\nu\rho} =
\partial_{\mu}\alpha_{\nu\rho}$. Since the Fr\o nsdal operator is
completely symmetric, the right-hand side has to be totally
symmetric as well. As it still has to satisfy the vanishing curl
condition (\ref{vanishcurl}), this is only possible if
$\alpha_{\nu\rho}=\partial_{\nu}\partial_{\rho}\alpha$, i.e.~if
 \bea\label{compensator}
  {\cal F}_{\mu\nu\rho} \ =\
  \partial_{\mu}\partial_{\nu}\partial_{\rho}\alpha\;.
 \eea
To conclude, the zero-curl equation (\ref{vanishcurl}) can be
integrated, giving rise to (\ref{compensator}). These are
precisely the compensator equations of Francia and Sagnotti
\cite{Francia:2002aa,Francia:2002pt}, with $\alpha$ being the
so-called compensator field. They are invariant under all
unconstrained HS transformations by virtue of the compensating
transformation
 \bea
  \delta_{\epsilon}\alpha\ =\ \epsilon^{\prime}\;.
 \eea
The appearance of the compensator should not come as a surprise
since we started with a HS Ricci tensor being invariant under
unconstrained HS transformations. To recover the constrained Fr\o
nsdal formulation one simply uses the fact that
$\epsilon^{\prime}$ acts as a shift symmetry on $\alpha$ in order
to set the latter to zero. Thus the dynamical content of
(\ref{vanishcurl}), in spite of being a 3$^{\rm rd}$ order
differential equation, is precisely that of a spin-3 mode on
Minkowski space.

The way we presented the derivation of (\ref{compensator}) was
entirely based on the `classical' Poincar\'e lemma. There is,
however, a more direct way thanks to the extended Poincar\'e lemma
of Dubois-Violette and Henneaux
\cite{DuboisViolette:1999rd,DuboisViolette:2001jk} (see also
\cite{Collins:1987bd,Bekaert:2002dt}), which applies to tensor
fields in more general Young tableaux than the completely
antisymmetric ones of differential forms. To be more precise, for
a sequence of Young diagrams with projectors $\mathbb{P}_p$ one
can define a differential $\tilde{d}$ as follows
 \bea
  \tilde{d} \ = \ \mathbb{P}_{p+1} \hspace{0.2cm}\circ\hspace{0.4em} \partial \;.
 \eea
It maps a tensor of degree $p$ to a tensor of degree $p+1$ by
first taking the derivative and then symmetrizing according to the
Young symmetries encoded by $\mathbb{P}_{p+1}$. If $N$ is the
maximal number of columns carried by the Young diagrams, we have
$\tilde{d}^{N+1}=0$. This reduces to the familiar $d^2=0$ for
differential forms, i.e.~for completely antisymmetric tableaux
with one column. In the case of a spin-3 field, equation
(\ref{vanishcurl}) can then schematically be written as
 \bea\label{closed}
  \tilde{d}{\cal F} \ = \ \big(\mathbb{P}_{\tiny \yng(3,1)}\hspace{0.2cm}
  \circ\hspace{0.4em}
  \partial\big){\cal F} \ = \ 0 \;.
 \eea
Here we have used the fact that the left-hand side of
(\ref{vanishcurl}) is in ${\tiny \yng(3,1)}$, but interpreted in
the antisymmetric basis, in agreement with the Bianchi identity
$({\rm Ric})_{[\mu\nu|\rho]\sigma}=0$. (See the discussion in
appendix C of \cite{Engquist:2007kz}.) Since the number of columns
is three, we have $\tilde{d}^4=0$, and so (\ref{closed}) implies
by the Poincar\'e lemma ${\cal F}=\tilde{d}^3\alpha$, whose
component form coincides with (\ref{compensator}). In this
language, the Fr\o nsdal operator $\cal F$ is a closed and
therefore exact form with respect to a generalized differential.

Let us now turn to the spin-4 case, which shows some new features.
The required top spin-4 connection determining the Riemann tensor
can be read off from (\ref{flatconn}). Its trace part, which
enters the Ricci tensor, can be written as
 \bea
  \begin{split}
   \omega_{\mu|\nu(3),\rho}{}^\s{}_\s
   \ =\ \partial_{\rho}\square h_{\nu(3)\mu}&-\partial_{(\nu_1}\square
   h_{\nu_2\nu_3)\rho\mu}+2\partial_{(\nu_1}\partial_{\nu_2}\partial\cdot
   h_{\nu_3)\rho\mu}\\
   &-2\partial_{\rho}\partial_{(\nu_1}\partial\cdot
   h_{\nu_2\nu_3)\mu}+\partial_{\rho}\partial_{(\nu_1}\partial_{\nu_2}h^{\prime}_{\nu_3)\mu}
   -\partial^3_{\nu(3)}h^{\prime}_{\rho\mu}\;.
  \end{split}
 \eea
Upon insertion this yields
 \bea\label{spin4ric}
  ({\rm Ric})_{\mu\nu|\rho(3),\sigma}\ =\ 4
  \del\undersym{{}_{\m}\del_{[\s}\cF_{(\r_1]\r_2\r_3)}{}_{\n}\hspace{.1cm}}
  \ =\ 0\;,
 \eea
where it is understood that the symmetrization has to be performed
at the very end. This can again be solved more conveniently by use
of the calculus of Dubois-Violette--Henneaux. In this language
(\ref{spin4ric}) is equivalent to
 \bea
  \tilde{d}^2{\cal F} \ = \ 0\;,
 \eea
where the sequence of Young diagrams consists of ${\tiny
\yng(4,1)}$ and ${\tiny \yng(4,2)}$. We have $\tilde{d}^5=0$ and
so by the Poincar\'e lemma ${\cal F}=\tilde{d}^3\alpha$. This is
precisely the compensator equation for spin-4,
 \bea\label{spin4comp}
  {\cal F}_{\mu\nu\rho\sigma}\ =\
  3\partial_{(\mu}\partial_{\nu}\partial_{\rho}\alpha_{\sigma)}\;.
 \eea
As before, this equation is invariant under unconstrained HS
transformations by virtue of the shift symmetry
$\delta_{\epsilon}\alpha_{\mu}=\epsilon^{\prime}_{\mu}$, which in
turn can be used to set the compensator to zero. But, there is
actually a novelty as compared to the spin-3 case, since all
fields with spin $s\geq 4$ can have a double trace part
$h^{\prime\prime}_{\mu_4\cdots\mu_s}$. The latter enters as
follows: Taking the divergence of the Fr\o nsdal operator and its
trace, say in the spin-4 case, results in the Bianchi identity
\cite{Francia:2002pt}
 \bea\label{bianchiF}
  \partial\cdot{\cal F}_{\mu\nu\rho}-\ft32 \partial_{(\mu}{\cal
  F}_{\nu\rho)}^{\prime} \ =\
  -\ft32\partial_{\mu}\partial_{\nu}\partial_{\rho}h^{\prime\prime}\;.
 \eea
This in turn implies that (\ref{spin4comp}) yields as a
consistency condition the additional equation
 \bea
  h^{\prime\prime} \ =\  \partial\cdot \alpha\;.
 \eea
Therefore, due to the Bianchi identity \eq{bianchiF}, gauge-fixing
$\alpha_{\mu}$ to zero, sets at the same time the double-trace
part of $h$ to zero. Thus one recovers indeed the Fr\o nsdal
equations together with the constraints (\ref{constraint0}), which
was required to consistently describe the spin-4 field.

In the remainder of this section we will turn to the general
spin-$s$ case. The Damour-Deser identity, which was crucial for
the above derivation, generalizes as follows: The frame-like
spin-$s$ Ricci tensor is given by
 \begin{eqnarray} \la{DDflat}
  ({\rm Ric})_{\mu\nu|\rho(s-1),\sigma(s-3)}&=&\ \frac{2s}{3}\mathbb
  P_{(s-1,s-3)}\del\undersym{{}_{\m}\del_{\s(s-3)}\cF_{\r(s-1)}{}_{\n}\hspace{.1cm}}
  \\ \nonumber
  &=& 2^{s-2}\mathbb{P}_{\rho(s-1)}\mathbb{P}_{\sigma(s-3)}\partial_{[\mu}
  \partial_{[\sigma_1}\cdots\partial_{[\sigma_{s-3}}{\cal
  F}_{\nu]\rho_1]\cdots\rho_{s-3}]\rho_{s-2}\rho_{s-1}}\;,
 \end{eqnarray}
where in the second line we have written out the projector
explicitly in order to make the curl-like structure manifest. It
is understood that we perform the antisymmetrizations over
$(\mu,\nu)$, $(\sigma_1,\rho_1)$, etc. This identity can be proven
as follows: We first notice that it has the $(s-1,s-3)$ Young
symmetry required by the left-hand side. Second, it is an HS
invariant $s$-th derivative object in the physical spin-$s$ field,
since $\delta_{\epsilon}{\cal
F}_{\mu_1\cdots\mu_s}\sim\partial_{(\mu_1}\partial_{\mu_2}
\partial_{\mu_3}\epsilon_{\mu_4\cdots\mu_s)}^{\prime}$.
So up to a factor, there is nothing else one can write down.
Finally the factor
originates from the $s-2$ antisymmetrizations. Using this
identity, the HS Einstein equation can be integrated as above to
the Fr\o nsdal equations in the compensator formulation, which
upon gauge-fixing sets the double-trace part to zero. We finally
note that the presented derivation of the Fr\o nsdal equations
coincides with the one given in \cite{Bekaert:2003az}, despite the
fact that we were starting in a frame-like formalism. 

\scss{Generalized Einstein equations on AdS} Let us now turn to
the HS dynamics on an AdS background. The unconstrained {\it
frame-like} formulation has so far been discussed only in case of
a spin-3 field in \cite{Engquist:2007kz}. In harmony with the
above derivation in flat space, the spin-3 case can be analyzed as
follows. Inserting (\ref{t2solu}) into the Ricci tensor and making
repeated use of the defining AdS relation for covariant
derivatives,
 \bea \la{nablaV}
  [\nabla_{\mu},\nabla_{\nu}] V_{\rho}\ =\
  \Lambda\left(g_{\nu\rho}V_{\mu}-g_{\mu\rho}V_{\nu}\right)\;,
 \eea
one finds that the Einstein equations can be written as
 \bea\label{adsspin3}
  ({\rm Ric})_{\mu\nu|\rho\sigma} \ =\
  R_{\mu\nu|\rho\sigma,\hspace{0.5em}\lambda}^{\hspace{2.1em}\lambda}
  \ =\ 2\nabla_{[\mu}^{}{\cal F}^{\rm AdS}_{\nu]\rho\sigma}\ =\ 0 \;.
 \eea
Thus we see that the Damour-Deser identity directly generalizes to
a spin-3 field on AdS. In principle, to systematically solve this
equation would require a refinement of the cohomological analysis
of Dubois-Violette and Henneaux to AdS. Here we will rather follow
a more pragmatic route by simply showing that it can be locally
solved in the required way. Indeed, one finds that
 \bea\label{sol}
  {\cal F}_{\mu\nu\rho}^{\rm AdS} \ =\
  \nabla_{(\mu}\nabla_{\nu}\nabla_{\rho)}\alpha-4\L
  g_{(\mu\nu}\nabla_{\rho)}\alpha\;
 \eea
solves (\ref{adsspin3}). Note that this is a solution only due to
the term proportional to the cosmological constant. The appearance
of this additional term comes as no surprise, since the right-hand
side has to take exactly the compensator form
\cite{Francia:2002pt}, which in turn is fixed by gauge invariance:
In fact, as above we have to set
$\delta_{\epsilon}\alpha=\epsilon^{\prime}$ in order for
(\ref{sol}) to be invariant under the unconstrained HS symmetry.

Next we turn to the spin-4 case, which is slightly more involved
due to the following reason. In sec.~\ref{torsection} we have seen
that the HS connections on AdS depend on the cosmological
constant, whose powers increase with the spin, see
eq.~(\ref{omeads9}). For a spin-4 field this means that the top
connection depends linearly on $\Lambda$, which upon insertion
gives rise to a Ricci tensor with a quadratic dependence. This in
turn implies that the naive AdS covariantization of the
Damour-Deser identity cannot be correct, since it depends, through
the mass-like term, only linearly on $\Lambda$. Therefore, the
Damour-Deser identity receives a correction, which is proportional
to $\Lambda$. One may compute this by explicit evaluation, but
this becomes rather tedious. Fortunately, it is possible to fix
this form by general reasoning as follows: Since the leading term
proportional to $\nabla^2{\cal F}$ is separately invariant under
constrained HS transformations with $\epsilon^{\prime}=0$, it
follows that the correction term has to be invariant as well.
Thus, since it is 2$^{\rm nd}$ order in derivatives (as one can
see from \eq{omeads9}), it has to be the Fr\o nsdal operator.
Finally, it remains to determine the relative coefficient. This
can be done by requiring gauge invariance or by using the fact
that the left-hand side is known to satisfy the Bianchi identity
(\ref{diffbianchi}). One finds
 \bea \la{x1x}
  ({\rm Ric})_{\mu\nu|\rho(3),\sigma}=\ \frac{8}{3}\mathbb
  P_{(3,1)}\left(\nabla\undersym{{}_{\m}\nabla_{\s}\cF_{\r(3)}{}_{\n}\hspace{.1cm}}
  -\Lambda g\undersym{{}_{\mu\sigma}{\cal F}_{\rho(3)}{}_{\nu}\hspace{.1cm}}\right)\;.
 \eea
In case of general spin we find the following expression
 \bea\la{DDAdS}
    \begin{split}
      (\rm Ric)_{\mu\nu|\rho(s-1),\sigma(s-3)}& \ =\ \frac{2s}{3}\mathbb
    P_{(s-1,s-3)}\Big(\nabla\undersym{{}_{\m}\nabla_{\s_1}\cdots\nabla_{\s_{s-3}}\cF_{\r(s-1)}{}_{\n}\hspace{.1cm}}\\
      &+\sum_{k=1}^{[\ft{s-2}{2}]}\vartheta^s_k\L^k g_{\s_{1}\s_2}\cdots
      g_{\s_{2k-3}\s_{2k-2}}g\undersym{{}_{\m\s_{2k-1}}\nabla_{\s_{2k}}\cdots\nabla_{\s_{s-3}}
      \cF_{\r(s-1)}{}_{\n}\hspace{.1cm}}
      \Big)\ .
      \end{split}
\eea
where $\vartheta^s_k$ are coefficients determined in appendix
\ref{appendixDD} to be
\bea \la{varthetasol}
    \vartheta^s_k\ = \ (-1)^k\t_{2k-1}\frac{(s-3)!}{(s-2k-2)!}\ ,
\eea
with $\t_{2k-1}$ the Taylor coefficients in $\tan x
=\sum_{k=1}^\infty \t_{2k-1} x^{2k-1}$.

After we have shown that the Ricci tensor satisfies a Damour-Deser
identity for general spin also on AdS, we are now able to recover
the Fr\o nsdal formulation. Since the resulting expression for the
Ricci tensor is gauge invariant, we can immediately conclude --
even without inspecting the precise coefficients -- that the
compensator ansatz
 \bea
  {\cal F}_{\mu_1\cdots\mu_s} \ =\ \ft12 (s-1)(s-2)\big(
  \nabla_{(\mu_1}\nabla_{\mu_2}\nabla_{\mu_3}\alpha_{\mu_4\cdots\mu_s)}
  -4\L g_{(\mu_1\mu_2}\nabla_{\mu_3}\alpha_{\mu_4\cdots\mu_s)}\big)\;
 \eea
solves the HS Einstein equation (\ref{HSEinstein}), as the
right-hand side takes precisely the form of the gauge variation of
${\cal F}$ under $\epsilon^{\prime}$, see \eq{gvoF}.

\scss{Towards an action principle?}\label{action} In this section
we will briefly discuss the possibility of a geometrical action
principle which gives rise to the Einstein equations
(\ref{HSEinstein}) in the frame formalism.\footnote{We are
grateful to Misha Vasiliev for discussions on this point.} More
precisely, we are asking if there is an action with the following
properties:
 \begin{itemize}
  \item[$(i)$]
   It should be written entirely in terms of differential forms, in
   analogy to the vielbein formulation of general relativity.
  \item[$(ii)$]
   It ought to be manifestly invariant under all (unconstrained)
   HS transformations.
  \item[$(iii)$]
   It should imply the torsion constraints together with
   the Einstein equations in a 1$^{\rm st}$ order formalism.
  \item[$(iv)$]
   Finally it should be local and not contain inverse powers of
   differential operators.\footnote{For non-local actions in the
   context of unconstrained frame fields see \cite{Bekaert:2006ix}.}
 \end{itemize}
Ideally one would like to satisfy requirement $(iii)$ in a way
that allows a 1.5 order formalism. This would mean that the
torsion constraint determining, say, $\omega^{a(s-1),b(t)}$ should
result from varying precisely this field, while the Einstein-like
equation should be obtained by varying with respect to the
physical HS field $e^{a(s-1)}$. In turn this would allow to
consistently eliminate the HS connections at the level of the
action, without altering the HS invariance or the Einstein
equations, simply due to the fact that any additional variation of
the HS connections resulting from this elimination would be
proportional to the vanishing torsion. Unfortunately, such a
formulation cannot exist, since the Einstein equations for spin
$s$ have $2s-2$ free indices in contrast to the $s$ indices of the
physical field to be varied (which coincide only for $s=2$). So
the best one can hope for is a formulation which yields the
required equations by varying with respect to some other fields.

In order to proceed we first note that condition $(i)$ in
combination with $(ii)$ implies that the action has to be written
entirely in terms of the Lorentz covariant HS field strengths
defined in sec.~\ref{HSalg}. To be specific, let us discuss the case
of a spin-3 field on a $D=4$ Minkowski space, which we expect to
exhibit generic features. We have to construct a 4-form out of the
HS curvatures, starting with $T^{ab}$. A term like $\bar{e}^a\wedge
\bar{e}^b\wedge T_{ab}$ vanishes identically due to the symmetry of
$T^{ab}$.
So the only possible terms seem to be those that contract HS
curvatures among themselves. Defining the Ricci 2-form
$R_{ab}=\ft12({\rm Ric})_{\mu\nu|ab}dx^{\mu}\wedge dx^{\nu}$ one
can write
 \bea\label{HSaction}
  S_{\rm HS} \ = \ \int T^{ab}\wedge T_{ab} + T^{ab,c}\wedge
  T_{ab,c} + T^{ab}\wedge R_{ab} \;,
 \eea
which is manifestly HS invariant. Up to the relative coefficients,
this is the unique expression, which is quadratic in spin-3
curvatures and carries trace parts only on the Riemann tensor. It
turns out that the dynamical content of (\ref{HSaction}) is to a
large extent independent on the precise value of these
coefficients. The equations of motion obtained from
(\ref{HSaction}) by varying with respect to $e^{ab}$,
$\omega^{ab,c}$ and $\omega^{ab,cd}$, respectively, read
 \begin{eqnarray}\label{eom1}
  dT^{ab} &=& 0\;, \\ \label{eom2}
  2T^{\langle ab}\wedge\bar{e}^{c\rangle}-2dT^{ab,c}+R^{\langle
  ab}\wedge \bar{e}^{c\rangle} &=& 0\;, \\ \label{eom3}
  2T^{\langle ab,c}\wedge \bar{e}^{d\rangle}-dT^{\langle
  ab}\eta^{cd\rangle} &=& 0\;.
 \end{eqnarray}
Inserting (\ref{eom1}) into (\ref{eom3}) implies
 \bea
  T^{\langle ab,c}\wedge \bar{e}^{d\rangle} \ = \ 0
  \hspace{1cm} \Longleftrightarrow \hspace{1cm} T^{ab,c} \ = \ 0\;,
 \eea
i.e.~we correctly recover one of the torsion constraints.
Unfortunately, for the other torsion constraint the equations of
motion (\ref{eom1}) imply only the weaker condition
$T^{ab}=d\xi^{ab}$, i.e.~the torsion tensor is exact, but not
necessarily zero. However, focusing on the particular solution
where it is zero yields with (\ref{eom2}) the Einstein equations
(\ref{HSEinstein}). We conclude that, while it seems to be
impossible along these lines to construct a manifestly HS
invariant action which satisfies the above requirements
$(i)$--$(iv)$, it is nevertheless possible to define an action
which at least contains a consistently propagating spin-3 mode,
potentially with dynamical torsion.

\scs{Conclusions and discussion} In this paper we presented a
systematic analysis of the frame or vielbein formulation of
unconstrained HS fields. We determined the torsion and Riemann
curvature tensor for arbitrary spin-$s$ fields. Imposing vanishing
torsion allowed us to express the HS connections in terms of the
physical field, whose solutions were determined explicitly in flat
space and AdS. The corresponding HS Christoffel symbols related
via a generalized metricity condition have been discussed. We
would like to stress that these are not identical to the
de\hspace{0.1cm}Wit--Freedman connections, though they are of
course related, as we have seen in sec.~\ref{metricity}. (So,
referring to the generalized spin-connections like in
\cite{Sagnotti:2005ns,Engquist:2007kz} as
de\hspace{0.1em}Wit-Freedman connections is slightly misleading.)
However, the de\hspace{0.1cm}Wit--Freedman connections are, in
contrast, not known on AdS in a closed form and have been
determined perturbatively in the inverse AdS length only recently
\cite{Manvelyan:2007hv}. Finally, we analyzed the HS dynamics in
this unconstrained formulation. As previously seen in the
metric-like formulation in flat space
\cite{Bekaert:2003az,Sagnotti:2005ns}, we found that the
higher-derivative Einstein equations satisfy the Damour-Deser
identity and can therefore be locally integrated to the Fr\o nsdal
equations in the compensator formulation. Moreover, we derived a
generalization of the Damour-Deser identity for AdS, which in turn
showed that also the AdS-Fr\o nsdal equations can be encoded in
the $s$-derivative Einstein equations. This confirms in particular
that the appearance of `extra fields' at the free-field level of
the Chern-Simons action in \cite{Engquist:2007kz} is not in
conflict with the requirement of standard physical field equations
of 2$^{\rm nd}$ derivative order. In other words, it verifies that
the higher derivatives, which naturally appear in HS theories, are
gauge artefacts.

We close with a few general comments on the advantages of this
unconstrained formalism, but also of its disadvantages, as
compared to the original one of \cite{Vasiliev:1980as}. First of
all, it is attractive since it parallels very closely the spin-2
case of gravity in vielbein form, as there is a clear distinction
between torsion and curvature tensors, while the latter gives rise
to manifestly HS invariant Einstein equations. Apart from that,
since we are ultimately interested in the coupling of HS fields to
gravity, the trace constraints that are inevitable in the
conventional frame formulation seem to be unnatural. Finally,
since the higher derivatives do appear anyway at the interacting
level of the HS theories of \cite{Fradkin:1986qy,Vasiliev:2001wa},
it seems to be legitimate to admit them also at the free-field
level, especially since they allow to nicely recover the Fr\o
nsdal formulation.

However, there are also disadvantages if one wants to encode the
dynamics in an action principle instead of equations of motion.
Specifically, as we discussed in sec.~\ref{action}, due to the
mismatch of free indices it is impossible to construct an action
which yields the Einstein equations by varying with respect to the
physical HS field. Instead, these field equations are obtained by
varying with respect to one of the HS connections. Moreover, it
seems to be difficult, if not impossible, to obtain all torsion
constraints from an action. These are definitely shortcomings,
since if one believes that only the HS field $h_{\mu_1\cdots
\mu_s}$ is of physical significance, there should be an action
principle formulated in terms of this field only. This is in
contrast to the constrained formulation of \cite{Vasiliev:1980as}.
For this, a 1$^{\rm st}$ order action can be given that contains
only a single connection, which in turn can be eliminated by its
equations of motion, giving rise to the standard 2$^{\rm nd}$
order Fr\o nsdal action. However, its HS invariance is not
manifest. In particular, without the presence of the corresponding
gauge field, the single connection is subject to one of the
St\"uckelberg shift symmetries discussed in sec.~\ref{HSalg},
which is possible only due to the tracelessness conditions.
Altogether this can be interpreted in the sense that, if one
insists on a manifestly HS invariant, unconstrained action
principle, the hierarchy of HS connections are not mere auxiliary
fields, but instead have to carry their own dynamics. In other
words, this indicates a propagating torsion as in the Chern-Simons
theories of \cite{Engquist:2007kz}. While from this point of view
there are certain unconventional features of an unconstrained
formalism, we think that it nevertheless possesses a number of
attractive properties which deserve further investigations.

\subsection*{Acknowledgments}
For useful comments and discussions we would like to thank
N.~Boulanger, P.~Sundell and M.A.~Vasiliev.

This work has been supported by the European Union RTN network
MRTN-CT-2004-005104 {\it Constituents, Fundamental Forces and
Symmetries of the Universe} and the INTAS contract 03-51-6346 {\it
Strings, branes and higher-spin fields}. O.H. is supported by the
stichting FOM.

\begin{appendix}
\renewcommand{\theequation}{\Alph{section}.\arabic{equation}}

\section{Notation and conventions}

Throughout the article, we work in $D$ dimensions so that the
space-time indices run between $\m,\n,\ldots=0,1,\ldots,D-1$.
Sometimes we utilize form language and use for a $p$-form
$F_p=\ft{1}{p!} F_{\m_1\cdots\m_p}dx^{\m_1}\wedge\cdots\wedge
dx^{\m_p}$. $A=0,1,\ldots,D-1,0^{\prime}$ denote $SO(D-1,2)$ vector
indices, out of which the first $D$ indices $a=0,1,\ldots,D-1$ are
Lorentz indices.

We use in the main text the language of Young tableaux, for which
our conventions are as follows: In case of AdS or Lorentz tensors
they encode the irreducible representations of $GL(D+1)$ or
$GL(D)$, respectively, since we are working with trace-full
tensors. We employ the symmetric basis with the convention that to
impose the Young-tableau symmetry of a diagram like

\vspace{-2cm}

  \setlength{\unitlength}{0.8cm}
  \begin{picture}(10,5)
  \put(6,0.5){\yng(8 ,5,3)}\put(6.55,-0.13){$\vdots$}\put(6,-1.5){\yng(2,1)}
  \put(10.9,1.85){$n_1$} \put(9.15,1.2){$n_2$} \put(6.75,-1.37){$n_r$}
  \end{picture} \vspace{1cm} \

\vspace{.2cm}

\noindent we first antisymmetrize over the columns and then
symmetrize the row indices. A Young diagram with $n_k$ boxes in
row $k$ is denoted by $(n_1,n_2,\ldots,n_r)$. Specifically, in
order to impose these symmetries on a tensor $X_{a^1_1\cdots
a^1_{n_1}|a^2_1\cdots a^2_{n_2}|\cdots|a^r_1\cdots a^r_{n_r}}$, we
first antisymmetrize in $a^1_k a^2_k\cdots a^{r}_k$ for
$k=1,\ldots,n_r$, and then symmetrize in $a^k_1\cdots a^k_{n_k}$
for $k=1,\ldots,r$.\footnote{We use a different convention to
impose Young-tableau symmetry than in \cite{Engquist:2007kz}.} In
particular, the action of a 2-row $(m,n)$ Young projector $\mathbb
P_{(m,n)}$ on a tensor $X_{a_1\cdots a_m|b_1\cdots b_n}$ with no a
priori symmetries in the two sets of indices reads ($n\le m$) \bea
\begin{split}
\mathbb P_{(m,n)}X_{a_1\cdots a_m|b_1\cdots b_n}  & \ = \
X_{\langle
a_1\cdots a_m|b_1\cdots b_n\rangle} \\
 & \ = \ \frac{m-n+1}{m+1}\mathbb P_{(m)}\mathbb
P_{(n)}\Big(X_{a_1\cdots a_m|b_1\cdots b_n}-X_{b_1a_2\cdots
a_m|a_1b_2\cdots b_n}\\ & \quad \qquad \qquad \qquad \qquad \qquad
+\cdots+(-1)^nX_{b_1\cdots b_na_1\cdots a_m|a_1\cdots a_n}\Big)\ ,
\end{split}
\eea
where $\mathbb P_{(m)}$ and $\mathbb P_{(n)}$ imposes total
(unit-weight) symmetrization in the two sets of indices. Here the
overall normalization has been chosen such that the projector
satisfies $\mathbb{P}^2=\mathbb{P}$.

\section{Recursion relations for AdS connections}\la{appendixRec}
In this appendix we derive the expression \eq{omeads9} for the HS
connections in terms of the physical fields in AdS. We are working
in the gauge-fixed formalism described in section
\ref{torsection}.

We start by recalling that the solution of
$\om_{\m|\n(s-1),\r(t)}$ in terms of $\O_{\r_t\m|\n(s-1),\r(t-1)}$
in \eq{Omgsol} is valid also in AdS since the identities
\eq{omegid} still hold after gauge fixing, as proven in section
\ref{sectionbianchi}. For $t=1$ the latter may be expressed in
terms of the physical field $h_{\m(s)}$ and we immediately deduce
the expression for the $t=1$ connection:
\bea \la{t1c}
    \om_{\m|\n(s-1),\r}&=&-\frac{s}{s-1}\O_{\m\langle\r|\n(s-1)\rangle}\ = \
    \frac{s}{s-1}\nabla_{\langle\r}h_{\n(s-1)\rangle\m}\ ,
\eea
whose explicit form is given in \eq{eqxx}. To determine the $t=2$
connection we first have to express $\O_{\r_2\m|\n(s-1),\r_1}$ in
terms of the physical field by using \eq{omcdab1} and the
expression for the $t=1$ connection just found, and then to insert
this into \eq{Omgsol} with $t=2$. The result for $\O$ is
\bea \la{zw3}
    \O_{\r_2\m|\n(s-1),\r_1}\ =\ \mathbb P_{(s-1,1)}\Big(\frac{s}{s-1}\nabla_{\r_2}\nabla_{\r_1}h_{\m\n(s-1)}
    +\L s ~h_{\n(s-1)\r_2}g_{\m\r_1}\Big)-(\r_2\leftrightarrow\m)\
    .
\eea
By noting that the terms in \eq{zw3} involving $h_{\n(s-1)\r_2}$
vanish identically under the $\mathbb P_{(s-1,2)}$ projector, we
recover the expression in \eq{t2solu}.

By using the formulas \eq{omcdab1} and \eq{Omgsol} one can easily
show that in general (with $1\le t\le s-1$) we have that
 \bea
     \nn\om_{\m|\n(s-1),\r(t)}&=&\frac{s-t+1}{s-t}\Big(\nabla_{\langle\r_t}\om_{\m|\n(s-1),\r(t-1)\rangle}\\
     &&  -\L(t-1)(s-t+2)\om_{\m|\langle\n(s-1),\r(t-2)}g_{\r_{t-1}\r_t\rangle}\Big)-(\r_t\leftrightarrow\m)\ . \la{tsc}
\eea Here we have used that the $(s-1,t-1)$ projector can be
suppressed under the $(s-1,t)$ projector. To solve for
$\om_{\m|\n(s-1),\r(t)}$ in terms of the physical field $h$, we
solve this equation iteratively by inserting the expressions for
$\om_{\m|\n(s-1),\r(t-1)}(h)$ and $\om_{\m|\n(s-1),\r(t-2)}(h)$
which are assumed to have been solved for in previous iteration
steps. (We remark that the $(\r_t\leftrightarrow\m)$ terms in
\eq{tsc} vanish under the projector $\mathbb P_{(s-1,t)}$ after
gauge-fixing $h$ to the completely symmetric part, since then the
$\omega$ are in irreducible $(s,t)$ tableaux.) For instance, the
$t=3$ connection can be solved for by inserting the expressions
for the $t=1$ and $t=2$ connections which are displayed in
\eq{eqxx} and \eq{t2solu}, respectively. Hence, this defines a
recursive problem which is solvable by induction.

The general solution is found to be given by
\bea \la{omeads10}
     \om_{\m|\n(s-1),\r(t)}& = &
     \sum_{k=0}^{[t/2]}\L^k\c_{k,t}g_{\langle\r_1\r_2}\cdots
     g_{\r_{2k-1}\r_{2k}}\nabla_{\r_{2k+1}}\cdots\nabla_{\r_t}h_{\n(s-1)\rangle\m}\ ,
\eea
where we have included the leading term in \eq{omeads9} as the
$k=0$ contribution. The coefficients $\c_{k,t}$ with $0\le k\le
[t/2]$ are determined by the recursive relation
\bea \la{xfr}
    \c_{k,t}& = & \frac{s-t+1}{s-t}\Big(\c_{k,t-1}-(t-1)(s-t+2)\c_{k-1,t-2}\Big)\ ,
\eea
with the `initial conditions'
\bea
 \c_{0,0}&=&1\ , \qquad \c_{1,2}\ = -\frac{s(s-1)}{s-2}\ .
\eea
Here it is understood that $\gamma_{k,t}=0$ for $t<0$.
Eq.~\eq{xfr} can be derived by inserting the expression
\eq{omeads10} into both sides of \eq{tsc} and comparing the
corresponding powers of $\L$. The explicit form of the first
coefficients is
 \bea
    \c_{0,t}&=& \frac{s}{s-t}\ , \\
    \c_{1,t}&=& \frac{st(t-1)\big(2t-1-3s\big)}{6(s-t)}\ , \\
    \c_{2,t}&=&\frac{st (t-1)(t-2)(t-3)
    \big(45 s^2-60 (t-1) s+4 t
    (5 t-12)+7\big)}{360 (s-t)}\ .
\eea
For even $t\ge4$ we find an expression for the $k=t/2$ coefficient
\bea
    \c_{t/2,t}&=&(-1)^{t/2}\frac{(t-1)(t-3)(s-t-1)(s-t+3)(s-t+4)}{s-t}\ .
\eea

\section{Damour-Deser identity in AdS}\la{appendixDD}

In this appendix we will derive the Damour-Deser identity
\eq{DDAdS} in AdS. The result follows from two requirements: $(i)$
that the relation involves an expansion in powers of $\L$ over
terms which have a linear dependence on the Fr\o nsdal operator
(see the discussion above \eq{x1x}); and $(ii)$ that the Bianchi
identity \bea
    \la{RicBI} \nabla_{[\l}({\rm Ric})_{\mu\nu]|\rho(s-1),\sigma(s-3)}\ = \ 0
\eea following from \eq{diffbianchi} holds. Note that $(i)$
implies that the identity reduces to the flat-space relation
\eq{DDflat} for $\L \rightarrow0$.

The requirement $(i)$ above implies that the expansion is given by
a sum over terms of the form $(\L g)^m \na^n \cF$, where $g$ is
the metric. These terms have to be compatible with dimensional
analysis and Young-tableau symmetry. Thus, the most general ansatz
which is compatible with the requirements $(i)$ and $(ii)$ above
takes the form
%
 \bea\la{Wansatz}
    \begin{split}
      (\rm Ric)_{\mu\nu|\rho(s-1),\sigma(s-3)}& \ =\ \frac{2s}{3}\mathbb
    P_{(s-1,s-3)}\Big(\nabla\undersym{{}_{\m}\nabla_{\s_1}\cdots\nabla_{\s_{s-3}}\cF_{\r(s-1)}{}_{\n}\hspace{.1cm}}\\
      &+\sum_{k=1}^{[\ft{s-2}{2}]}\vartheta^s_k\L^k g_{\s_{1}\s_2}\cdots
      g_{\s_{2k-3}\s_{2k-2}}g\undersym{{}_{\m\s_{2k-1}}\nabla_{\s_{2k}}\cdots\nabla_{\s_{s-3}}
      \cF_{\r(s-1)}{}_{\n}\hspace{.1cm}}
      \Big)\ .
      \end{split}
\eea
The challenge is to determine the coefficients $\vartheta^s_k$.
Note that by this ansatz the algebraic Bianchi identity
(\ref{algbi}) is identically satisfied and does not lead to any
further constraint.

Let us illustrate the method by fixing the first two coefficients
$\vartheta^s_1$ and $\vartheta^s_2$. To simplify the presentation
we use the notation $\nabla_{\m_1\m_2\cdots\m_n}\equiv
\nabla_{\m_1}\nabla_{\m_2}\cdots \nabla_{\m_n}$, with the indices
arranged in this particular order. We focus on terms proportional
to $\L$ in (\ref{RicBI}), which are given by
 \bea\label{term1}
    \begin{split}
    &\mathbb P_{(s-1,s-3)}\Big([\nabla_\l,\nabla_\m]\nabla_{\s_1\cdots\s_{s-3}}\cF_{\r(s-1)\n}+
    [\nabla_\m,\nabla_\n]\nabla_{\s_1\cdots\s_{s-3}}\cF_{\r(s-1)\l}\\&\qquad\qquad\qquad+
    [\nabla_\n,\nabla_\l]\nabla_{\s_1\cdots\s_{s-3}}\cF_{\r(s-1)\m}
    +3!\vartheta^s_1\L g_{\s_{s-3}[\m}\nabla_{\l|\s(s-4)|}\cF_{\n]\r(s-1)}\Big)\ .
    \end{split}
\eea
In order to determine $\vartheta_1^s$ it is sufficient to inspect
terms of a specific structure, say,
$g_{\mu\sigma_{s-3}}\nabla_{\sigma(s-4)}$. Using \eq{nablaV},
these can be written as
\bea\label{term2}
    \L\mathbb P_{\r(s-1)}\mathbb P_{\s(s-3)}g_{\m\s_{s-3}}\Big(\big(\nabla_{\l\s(s-4)}+\cdots+\nabla_{\s(s-4)\l}\big)
    \cF_{\r(s-1)\n}+\vartheta^s_1
    \nabla_{\l\s(s-4)}\cF_{\r(s-1)\n}\Big)\ .
\eea
When performing the $(s-1,s-3)$ projection, we were allowed to
ignore the antisymmetrization contained in it, since they would
give rise to different index structures. Next we have to commute
the covariant derivatives such that they all appear in the form
$\nabla_{\l\s(s-4)}$. One finds
\bea\label{term3}
    \begin{split}
    &\L\mathbb P_{\r(s-1)}\mathbb
    P_{\s(s-3)}g_{\m\s_{s-3}}\Big((\vartheta^s_1+(s-3))
    \nabla_{\l\s(s-4)}\cF_{\r(s-1)\n}+(s-4)[\nabla_{\s_{s-4}},\nabla_\l]\na_{\s(s-5)}\\&\qquad\qquad\qquad
    +(s-5)\na_{\s_{s-4}}[\nabla_{\s_{s-5}},\nabla_\l]\na_{\s(s-6)}
    +\cdots+\na_{\s(s-5)}[\nabla_{\s_{s-4}},\nabla_\l]\Big)\cF_{\r(s-1)\n}\
    .
    \end{split}
 \eea
Therefore we conclude
 \bea
  \vartheta^s_1\ =\ -(s-3)\;,
 \eea
while the commutators give rise to terms that are quadratic in
$\L$ and therefore have to be cancelled by higher-order
contributions. As a consistency check we note that the terms
proportional to $g_{\m\r_{s-1}}$, $g_{\m\n}$ and $g_{\m\l}$ vanish
identically, which follows from the total symmetry of the Fr\o
nsdal operator.

Let us now turn to the next, i.e., quadratic order. For this we
evaluate the commutators in \eq{term3} and include the $k=2$ term
in \eq{Wansatz}, to arrive at an analogous expression to
\eq{term2}:
\bea\label{term4}
    \begin{split}
    -\L^2\mathbb P_{\r(s-1)}\mathbb
    P_{\s(s-3)}g_{\m\s_{s-3}}g_{\s_{s-5}\s_{s-4}}\Big(&\big(
    \chi_{s-4}\nabla_{\l\s(s-6)}
    +\cdots+\chi_{2}\nabla_{\s(s-6)\l}\big)\cF_{\r(s-1)\n}\\
    &+\vartheta^s_2
    \nabla_{\l\s(s-6)}\cF_{\r(s-1)\n}\Big) \ ,
    \end{split}
 \eea
where we have defined
\bea
    \chi_m\ : = \ \sum_{n=m}^{s-4}n\ = \ \ft12(s-3-m)(s-4+m)\ .
\eea
Then focusing on terms with the index structure
$g_{\mu\sigma}g_{\sigma\sigma}$ determines the next coefficient to
be
 \bea
  \vartheta^s_2 \ =\ \sum_{m=2}^{s-4}\chi_m\ = \ \ft13 (s-3)(s-4)(s-5)\;.
 \eea
The general case proceeds in exact analogy. By repeating the steps
just described and focusing on a specific index structure, we
ultimately obtain the simple form displayed in \eq{varthetasol}.
As a consistency check, we verified up to cubic order in $\L$ that
all other terms cancel.

\end{appendix}

\end{document}